\documentclass[aip, reprint, longbibliography, floatfix]{revtex4-2}

\usepackage[utf8]{inputenc}
\usepackage{multirow}

\usepackage[lining,semibold]{libertine}
\usepackage[libertine, cmintegrals, bigdelims, vvarbb]{newtxmath}

\usepackage{amsmath}
\usepackage{amsfonts}
\usepackage{mathrsfs}
\usepackage{gensymb}
\usepackage{bbm}
\usepackage{dsfont}

\usepackage{chemformula}
\usepackage[caption=false]{subfig}
\usepackage{chemfig}
\usepackage[version=3]{mhchem}

\usepackage{tikz}
\usetikzlibrary{matrix,positioning,decorations.pathreplacing,arrows.meta,calc,bending}

\usepackage{braket}

\usepackage{soul}
\usepackage{xcolor}

\usepackage{scalerel}
\usepackage{comment}

\usepackage{blkarray}

\usetikzlibrary{arrows, automata}
\usepackage{algorithm}
\usepackage{algpseudocode}

\definecolor{webgreen}{rgb}{0,.5,0}
\definecolor{webbrown}{rgb}{.6,0,0}
\definecolor{grigio}{rgb}{.85,.85,.85} 
\definecolor{RoyalBlue}{rgb}{0.0, 0.14, 0.4}
\definecolor{skyblue1}{rgb}{0.45,0.62,0.81}
\definecolor{skyblue2}{rgb}{0.2,0.39,0.64}
\definecolor{skyblue3}{rgb}{0.13,0.29,0.53}
\definecolor{scarlet1}{rgb}{0.93,0.16,0.16}
\definecolor{scarlet2}{rgb}{0.8,0,0}
\definecolor{scarlet3}{rgb}{0.64,0,0}

\definecolor{g}{gray}{0.50}

\usepackage{hyperref}
\hypersetup{%
    colorlinks=true, linktocpage=true, pdfstartpage=1, pdfstartview=FitV,%
    breaklinks=true, pdfpagemode=UseNone, pageanchor=true, pdfpagemode=UseOutlines,%
    plainpages=false, bookmarksnumbered, bookmarksopen=true, bookmarksopenlevel=1,%
    hypertexnames=true, pdfhighlight=/O,%
    urlcolor=webbrown, linkcolor=RoyalBlue, citecolor=webgreen,
    pdftitle={},%
    pdfauthor={Benedikt Remlein},%
    pdfsubject={},%
    pdfkeywords={},%
    pdfcreator={pdfLaTeX},%
    pdfproducer={LaTeX REVTeX}%
}

\usepackage[capitalise]{cleveref}

\usepackage{verbatim}

\usepackage{kbordermatrix}

\usetikzlibrary{calc}
\newcommand{\crn}[2]{\ch{<=>[$#2$][$#1$]}}

\newcommand{\figwidth}{0.95\linewidth}

\newcommand{\remark}[1]{%
\par\textit{#1}\quad
}

\makeatletter
\def\maketag@@@#1{\hbox{\m@th\normalfont\normalsize#1}}
\makeatother

\DeclareMathAlphabet{\mathpzc}{OT1}{pzc}{m}{it}

\begin{document}

    \title{A minimal thermodynamically consistent chemical oscillator}

    \author{Benedikt Remlein}
    \email{benedikt.remlein@uni.lu}
    \affiliation{Complex Systems and Statistical Mechanics, Department of Physics and Materials Science, University of Luxembourg, 30 Avenue des Hauts-Fourneaux, L-4362 Esch-sur-Alzette, Luxembourg}

    \author{Luca Monari}
    \email{l.monari@zmbh.uni-heidelberg.de}
    \affiliation{Institut de Science et d'Ingénierie Supramoléculaires (ISIS), University of Strasbourg, CNRS \& icFRC, UMR 7006, 8 Allée Gaspard Monge, 67000 Strasbourg, France}
    \affiliation{Present address: Biophysical Engineering Group, Center for Molecular Biology of Heidelberg University (ZMBH), Heidelberg University, Heidelberg, Germany}

    \author{Beatrice Bartolomei}
    \email{beatrice.bartolomei@northwestern.edu}
    \affiliation{Department of Chemical and Pharmaceutical Sciences, CENMAT, Centre of Excellence for Nanostructured Materials, INSTM UdR Trieste, University of Trieste, Via Licio Giorgieri 1, Trieste, 34127, Italy}
    \affiliation{Present address: Department of Chemistry, Northwestern University, Evanston, Illinois, 60208, USA}

    \author{Giulio Ragazzon}
    \email{ragazzon@unistra.fr}
    \affiliation{Institut de Science et d'Ingénierie Supramoléculaires (ISIS), University of Strasbourg, CNRS \& icFRC, UMR 7006, 8 Allée Gaspard Monge, 67000 Strasbourg, France}
    
    \author{Massimiliano Esposito}
    \email{massimiliano.esposito@uni.lu}
    \affiliation{Complex Systems and Statistical Mechanics, Department of Physics and Materials Science, University of Luxembourg, 30 Avenue des Hauts-Fourneaux, L-4362 Esch-sur-Alzette, Luxembourg}
        
    \author{Emanuele Penocchio}
    \email{penocchio@unistra.fr}
    \affiliation{Institut de Science et d'Ingénierie Supramoléculaires (ISIS), University of Strasbourg, CNRS \& icFRC, UMR 7006, 8 Allée Gaspard Monge, 67000 Strasbourg, France}
    
	\date{\today}

\begin{abstract}

Inspired by the chemical reaction network considered as the smallest system featuring a Hopf bifurcation and its reversible extension, we introduce an even more minimal reaction network that is thermodynamically consistent and exhibits autonomous oscillations under nonequilibrium driving. The model combines three features that are rarely realized simultaneously in compact oscillator networks: a chemically plausible structure restricted to uni- and bimolecular reactions, reversibility of such reactions, and analytical tractability. The system consists of three internal species coupled to two chemostatted species and still undergoes a supercritical Hopf bifurcation when a chemostat concentration is varied.
To analyze the dynamics and thermodynamics near the onset of oscillations, we employ the mathematical technique of normal-form reduction, which allows obtaining a controlled irreversible approximation that preserves the leading phase-space structure of the full reversible network while enabling explicit calculations. This framework provides analytical access to the bifurcation structure and the leading contributions governing thermodynamic observables.
Within this setting, we use the onset of oscillations to characterize the thermodynamic response of the system. In particular, we offer an analytical description of key thermodynamic quantities such as the semi-grand Gibbs free energy and the non-conservative work rate, which exhibit a kink-like discontinuity at the Hopf bifurcation. We thus provide an analytical description of a phenomenon previously characterized only through numerical simulations of more complex networks.

\end{abstract}

\maketitle 

\section{Introduction}

Autonomous chemical oscillations are a central motif in nonequilibrium chemical systems, underlying processes ranging from biochemical clocks and signaling cycles to synthetic reaction networks designed for autonomous function.\cite{epst96,gasp02a,gonz02,Falcke2004,selk68,Gentili20} Identifying compact reaction schemes\cite{Micheau92,wilh95,Banaji_2023} that realize such behavior while remaining chemically plausible is important to bridge mathematical descriptions with chemical realizations, as well as for testing physical theories without compromising the generality of findings; it thus has implications for systems chemistry in its broadest interdisciplinary interpretation.\cite{ashk17}
Experimental realizations include autocatalytic organic oscillators built from thiol–disulfide and small-molecule chemistries,\cite{terHarmsel2023,Semenov26} enzymatically and hybridization-fueled nucleic-acid networks,\cite{Kim2011,Montagne2011,Franco19} and chemically driven dissipative supramolecular systems exhibiting transient assembly and sustained oscillations.\cite{Boekhoven2015,Sorrenti2017,Leira2018,Boek25,Weber26} 

At the same time, recent developments in stochastic thermodynamics\cite{seif12,vand15,peli21,seifert} provide a consistent framework to quantify energetic costs in driven chemical reaction networks (CRNs), including work, dissipation, and generalized thermodynamic potentials.\cite{schm06,rao16,rao18,avan24} In this context, it was also shown numerically that oscillatory transitions in driven systems can leave signatures (e.g., a kink-like discontinuity) in thermodynamic observables.\cite{fala18,nguy18, herp18, meib24,gopa25} These developments motivated us to investigate oscillatory models that are chemically plausible, minimal and thermodynamically consistent, while allowing analytical tractability: a combination that remains essentially unexplored. Indeed, detailed kinetic descriptions of oscillatory systems are often high-dimensional (i.e., they involve many reactions), whereas widely used minimal oscillator models typically rely on effective schemes including higher-order reactions, or on irreversible steps that obscure a transparent thermodynamic interpretation.\cite{fiel74,fiel75,prig68,schn79,lefe88} 
However, for chemically plausible oscillators that include at most bi-molecular reactions a minimal reaction scheme contains at least three internal species.\cite{wilh95} A recent study systematically examined minimal irreversible reaction networks sharing these features,\cite{Banaji_2023} albeit lacking thermodynamic consistency. Earlier minimal oscillator constructions provide compact deterministic descriptions,\cite{wilh95} and reversible extensions have been considered,\cite{wilh97} but often require additional complexity or do not fully retain analytical accessibility.

In this work, we address the outlined limitation of the state of the art. Inspired by the reversible version of the smallest CRN with Hopf bifurcation\cite{wilh95,wilh97}, we introduce an even more minimalistic reversible CRN that exhibits a supercritical Hopf bifurcation under nonequilibrium driving. The network consists of three internal species, the minimum required for Hopf bifurcation, and two chemostatted species, the minimum required to sustain nonequilibrium driving, while involving only uni- and bimolecular reactions. This provides a chemically plausible and thermodynamically consistent realization of an autonomous oscillator in a minimal setting. We analyze the deterministic mass-action dynamics of the model and characterize its thermodynamic behavior in terms of its core thermodynamic quantities: the semi-grand Gibbs free energy, the non-conservative work rate, and the entropy production rate. To obtain analytical insight, we combine a normal-form reduction near the Hopf bifurcation\cite{guck83,craw91,wigg03} with a controlled irreversible approximation that preserves the leading phase-space structure while enabling explicit calculations. This approach allows us to determine the bifurcation features and the leading thermodynamic response in a compact chemically explicit model. As an application, we show that the onset of oscillations is accompanied by kink-like discontinuities in time-averaged thermodynamic observables. While in previous works their occurrence in more elaborated networks was observed from numerical simulations, here they are captured by the analytical description.\cite{fala18,nguy18, herp18, meib24,gopa25} 

Overall, the analytical approach developed here provides a direct route to understanding these features from the underlying reaction-network dynamics. In this respect, the model also connects to the broader program of designing out-of-equilibrium chemical functions through artificial molecular ratchets\cite{KayZerb2007,FeringaLeigh17,AC_MB_SS_2020,ragazzon2018,Das2021,Amano2021,Sangchai2023,Borsley2024,Penocchio2024} in which minimal CRN motifs serve as building blocks for autonomous chemical behavior.

This manuscript is organized as follows. We introduce the reversible CRN in Sec.~\ref{sec:Model} and discuss its thermodynamic description in Sec.~\ref{sec:TD}. Section~\ref{sec:Numerics} presents numerical evidence for the Hopf bifurcation and the associated thermodynamic response. In Sec.~\ref{sec:TDObservables}, we analyze the vicinity of the bifurcation using a mapping to the normal form of a Hopf bifurcation. Section~\ref{sec:Irreversible} develops the irreversible approximation and derives analytical estimates for the thermodynamic observables. We conclude in Sec.~\ref{sec:Conclusion}. Additional details are provided in the Appendices.

\section{Minimal CRN}
\label{sec:Model}

We propose a CRN that consists of the following reacting species $\mathcal S \equiv \{X,Y,Z,S,P\}$. The reacting species form a well-stirred mixture surrounded by a much more abundant non-reacting solvent species that acts as both a heat reservoir at constant temperature $T$ and a volume reservoir at constant volume $V$. 

The reacting species $\mathcal S$ are interconverted by the following chemical reactions,
\begin{equation}
    \begin{split}
    S+X&\crn{-1}{1}2X\,,\\
    X&\crn{-2}{2}Z\,,\\
    Z&\crn{-3}{3}Y\,,\\
    Y&\crn{-4}{4}P\,,\\
    X+Y&\crn{-5}{5}P+Y\,,
    \end{split}\label{eq:CRN}
\end{equation}
where $\rho \in \mathcal R \equiv \{\pm 1,\pm2,\pm3,\pm4,\pm5 \}$ labels a reaction and $-\rho \in \mathcal R$ its corresponding backward reaction. Thus, CRN (\ref{eq:CRN}) is reversible.

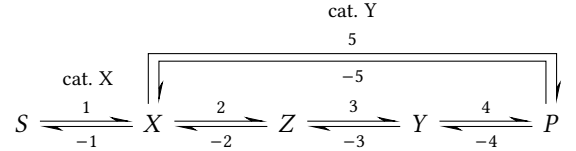
\begin{figure}[htbp]
\centering
\begin{tikzpicture}[
    node distance=13mm,
    species/.style={font=\normalsize},
    cat/.style={font=\footnotesize, inner sep=1.5pt},
    half/.style={-{Stealth[left, length=2.8mm, width=1.2mm]}, shorten >=1pt, shorten <=1pt},
    halfswap/.style={-{Stealth[left, length=2.8mm, width=1.2mm]}, shorten >=1pt, shorten <=1pt},
]
\node[species] (S) {$S$};
\node[species] (X) [right=of S] {$X$};
\node[species] (Z) [right=of X] {$Z$};
\node[species] (Y) [right=of Z] {$Y$};
\node[species] (P) [right=of Y] {$P$};

\newcommand{\rev}[4]{%
  \draw[half]     ([yshift= 1pt]#1.east) -- node[above=0pt,font=\scriptsize]{$#3$} ([yshift= 1pt]#2.west);
  \draw[halfswap] ([yshift=-1pt]#2.west) -- node[below=0pt,font=\scriptsize]{$#4$} ([yshift=-1pt]#1.east);
}

\rev{S}{X}{1}{-1}
\rev{X}{Z}{2}{-2}
\rev{Z}{Y}{3}{-3}
\rev{Y}{P}{4}{-4}

\node[cat] (c1) at ($(S)!0.5!(X)+(0,6mm)$) {cat. X};

\draw[half] ([xshift=-2pt]X.north) -- ++(0,7mm)
      -- node[above,font=\scriptsize]{$5$} ([xshift=2pt,yshift=7mm]P.north)
      -- ([xshift=2pt]P.north);
\draw[halfswap] ([xshift= -2pt]P.north) -- ++(0,6mm)
      -- node[below,font=\scriptsize]{$-5$} ([xshift= 2pt,yshift=6mm]X.north)
      -- ([xshift= 2pt]X.north);
\node[cat] at ($(X)!0.5!(P)+(0,15mm)$) {cat. Y};
\end{tikzpicture}
\caption{Schematics of the reaction network associated with the set of reactions in equation \eqref{eq:CRN}, with catalyzed steps explicitly indicated.}
\label{fig:CRN}
\end{figure}

The CRN structure can be appreciated by drawing it as in Fig.~\ref{fig:CRN}, which highlights that two paths are available to convert species $S$ (a notation selected to signify the role of a substrate) into species $P$ (selected to indicate the product). The reacting species are thus partitioned into internal species $\mathcal S_i \equiv \{X,Y,Z\}$ and chemostatted species $\mathcal S_c \equiv \{S,P\}$. The internal species are three as in the network proposed by Wilhelm \emph{et~al.},\cite{wilh97} to which this work is inspired, but here the chemostatted species are two instead of three\footnote{in the original network, the forward reactions 4 and 5 were leading to two different species, now combined into the single chemostat $P$}, offering an additional simplification. The concentrations of these two chemostats are kept constant through a not further specified exchange process with the environment of the CRN. This approach allows controlling the energy available to the system by imposing the concentration of chemostatted species. The internal species $X$ and $Y$ act also as catalysts. In particular, reaction 1 is autocatalytic in $X$ and is part of both paths that convert $S$ into $P$, while reaction 5 is catalyzed by $Y$, and is part of only one of the two paths. As a result, the relative rate at which $P$ is formed through a given path depends on the concentration of species $Y$.

Equipping CRN (\ref{eq:CRN}) with mass action dynamics \cite{gasp04}, its deterministic dynamics is governed by the following reaction rate equation (RRE),
\begin{equation}
    \frac{d}{dt}[\boldsymbol z]_t = \mathbb S\boldsymbol J([\boldsymbol z]_t) + \boldsymbol I_t\label{eq:RRE}
\end{equation}
where $[\boldsymbol z]_t \equiv (\ldots\,[\alpha_t]\,\ldots)^T$ denotes the vector of the concentrations of species $\alpha \in \mathcal S$ at time $t$,
and $\mathbb S$ the stoichiometric matrix,
\begin{equation}
   \mathbb S \equiv (\ldots\,\mathbb S_\rho\,\ldots) \equiv
 \kbordermatrix{
    & \color{g}1 &\color{g}2&\color{g}3&\color{g}4&\color{g}5\cr
    \color{g}X 	  &1& -1& 0& 0& -1\cr
    \color{g}Y  	  &0& 0& 1& -1& 0\cr
    \color{g}Z  	  &0& 1& -1& 0& 0\cr
    \color{g}S   &-1&0& 0& 0& 0\cr
    \color{g}P        &0& 0& 0& 1& 1\cr
  }\,,
\end{equation}
whose entries $\mathbb S_{\alpha,\rho}$ are indexed so that rows label species and columns label reactions (as indicated by the gray annotations). Accordingly, the column-vectors $\mathbb S_\rho = (\ldots\,\mathbb S_{\rho,\alpha}\,\ldots)^T$ encode the net-stoichiometric change $\mathbb S_{\rho,\alpha}$ of species $\alpha \in \mathcal S$ in reaction $\rho\in\mathcal R$. 
Furthermore, the vector of reaction-currents $\boldsymbol J([\boldsymbol z]_t)$ in Eq.~(\ref{eq:RRE}) is defined as
\begin{equation}
    \boldsymbol J([\boldsymbol z]_t) \equiv 
 \kbordermatrix{
 & \\
    \color{g}1 	  &k_1 [S][X]_t - k_{-1}[X]_t^2\\
    \color{g}2  	  &k_2 [X]_t - k_{-2}[Z]_t\\
    \color{g}3  	  &k_3 [Z]_t - k_{-3}[Y]_t\\
    \color{g}4   &k_4 [Y]_t - k_{-4}[P]\\
    \color{g}5        &k_5 [X]_t[Y]_t - k_{-5}[P][Y]_t
  }\,,
\end{equation}
with the kinetic constant $k_\rho$ of reaction $\rho \in \mathcal R$, and the vector of external currents $\boldsymbol I_t$, which models the effective chemostatting process, \cite{avan22,avan24} is given by
\begin{equation}\boldsymbol I_t \equiv 
 \kbordermatrix{
 & \\
    \color{g}X 	  &0\\
    \color{g}Y  	  &0\\
    \color{g}Z  	  &0\\
    \color{g}S   &I_S\\
    \color{g}P        &I_P
  }\,.
\end{equation}
The currents $I_S$ and $I_P$ are chosen such that the corresponding concentrations $[S]$ and $[P]$ remain constant in time.

\section{Thermodynamics of the minimal CRN}
\label{sec:TD}

Due to the exchange with the environment (modeled by the current $\boldsymbol I_t$), CRN (\ref{eq:CRN}) is said to be open. Here, we introduce the basic thermodynamic notation of open CRNs equipped with mass-action dynamics and discuss its application to the model introduced in Sec. \ref{sec:Model}. 
To that end, we assume that all reactions are elementary, i.e., all reactions and reacting species are resolved, or emerge from an effective, coarse-grained description that attributes to each reaction a unique reaction current. \cite{VanRysselberghe1958, avan23}

For such networks,
the local detailed balance condition (LDB) holds for every reaction $\rho \in \mathcal R$, \cite{rao18}
\begin{equation}
    RT \ln \frac{k_\rho}{k_{-\rho}} = - \boldsymbol \mu^\circ\cdot \mathbb S_\rho\,,\label{eq:LDB}
\end{equation}
where $\boldsymbol \mu^\circ \equiv (\ldots \mu_\alpha^\circ\ldots)$ denotes the vector of standard chemical potentials of the reacting species $\mathcal S$, and $R$ denotes the gas constant. The reversibility of CRN (\ref{eq:CRN}) together with Eq. (\ref{eq:LDB}) ensures a thermodynamically consistent description of the CRN \cite{schm06, rao16, rao18} within the ideal-dilute regime \--- the framework extends to non-ideal mixtures through activity-coefficient generalizations of local detailed balance.\cite{Avanzini2021} Thus, the Gibbs free energy is given by
\begin{equation}
G \equiv \sum_{\alpha \in \mathcal S} \big(\mu_\alpha([\alpha])  - RT\big)[\alpha]\,,
\end{equation}
where $\mu_\alpha([\alpha]) \equiv \mu_\alpha^\circ + RT \ln[\alpha]$ denotes the chemical potential of species $\alpha\in \mathcal S$ in the ideal solution approximation. For convenience, in the following, we set $RT = 1$.

The entropy production rate (in units of the gas constant $R$) is given by
\begin{equation}
    \dot \Sigma \equiv -\boldsymbol \mu\cdot\mathbb S\boldsymbol J\,,
\end{equation}
where $\boldsymbol \mu \equiv (\ldots\mu_\alpha([\alpha])\ldots)$ is the vector of chemical potentials, and $\boldsymbol a \cdot \boldsymbol b \equiv \sum_\alpha a_\alpha b_\alpha$ denotes the standard scalar product. The work rate to maintain the fixed concentrations of the chemostatted species reads
\begin{equation}
    \dot W_\mathrm{chem} \equiv \boldsymbol \mu \cdot \boldsymbol I\,.
\end{equation}
The previously defined thermodynamic quantities are related by a first-law like energy-balance equation,
\begin{equation}
   - \frac{d}{dt}G =  \dot \Sigma - \dot W_\mathrm{chem}\,.\label{eq:EnergyBalance}
\end{equation}

Relation (\ref{eq:EnergyBalance}) shows that the Gibbs free energy is generally not the appropriate thermodynamic potential for an open CRN. From the second law, $\dot \Sigma\geq 0$, and Eq.~(\ref{eq:EnergyBalance}), one gets 
$-\frac{d}{dt}G \geq - \dot W_\mathrm{chem}$,
such that the sign of $\dot G$ is not fixed by the second law. 
In particular, $G$ can be non-monotonic even when the chemostats are at equal chemical potential, $\mu_S=\mu_P$, since the chemical work rate $\dot W_\mathrm{chem}=\mu_S(I_S-I_P)$ generally remains nonzero away from the steady state. This reflects that $G$ contains the energetic contribution associated with the exchange of matter with the environment. To isolate the internal relaxation dynamics, we therefore subtract this contribution and define the semi-grand Gibbs free energy \cite{rao18,fala18,avan24}
\begin{equation}
    \mathcal G \equiv G - \mu_P([P]) [L]\,.
    \label{eq:SemigrandGibbs}
\end{equation}
Here, $[L] \equiv \sum_{\alpha\in \mathcal S} [\alpha]$ denotes the total concentration, which is no longer conserved in the open CRN (\ref{eq:CRN}). Subtracting the reference contribution $\mu_P([P])[L]$ removes the energetic contribution associated with the exchange of matter with the environment.

Taking the time-derivative of $\mathcal G$, we find a balance equation,
\begin{equation}
    -\frac{d}{dt}\mathcal G =  \dot \Sigma - \dot W\,,\label{eq:1stCurly}
\end{equation}
where 
\begin{equation}
    \dot W \equiv \big(\mu_S([S]) -\mu_P([P])\big)I_S\label{eq:WorkRate}
\end{equation} 
denotes the non-conservative work-rate, which quantifies the energetic cost to maintain the current $I_S$ through the network due to the chemostatting of species $S$ and $P$.
Using Eq. (\ref{eq:RRE}), the external current is given by $I_S = k_1 [S][X]_t-k_{-1} [X]_t^2$.

The decomposition in Eq.~(\ref{eq:1stCurly}) separates the free-energy dissipation from the externally imposed driving work. This property will prove useful in Sec.~\ref{sec:Irreversible}, where the reversible CRN~(\ref{eq:CRN}) is approximated by an effective irreversible CRN that captures the leading-order dynamics while retaining a consistent thermodynamic interpretation.

\remark{Remark (internal cycle):}
CRN (\ref{eq:CRN}) has an internal cycle $\mathcal C$, i.e., a series of reactions that, after completion, do not change the state of a closed CRN \cite{avan24}. The internal cycle is characterized by the vector $\boldsymbol c \equiv \begin{pmatrix}0&1&1&1&-1\end{pmatrix}^T$ with
\begin{equation}
    \mathbb S\boldsymbol c = \boldsymbol 0\,,
\end{equation}
which corresponds to the reactions $\mathcal C \equiv \{2,3,4,-5\}$, as one may appreciate by looking at Fig.~\ref{fig:CRN}. In a closed CRN, $\boldsymbol c$ describes the direction of a stationary reaction current, compare Eq. (\ref{eq:RRE}) for $\boldsymbol I = \boldsymbol 0$. The local detailed balance condition, Eq. (\ref{eq:LDB}), evaluated along the internal cycle $\mathcal C$ leads to the following condition on the kinetic constants,
\begin{equation}
	\frac{k_2k_3k_4k_{-5}}{k_{-2}k_{-3}k_{-4}k_5} = e^{-\sum\limits_{\rho \in \mathcal C} \boldsymbol \mu^\circ \cdot \boldsymbol S_\rho} = e ^0 = 1\,.\label{eq:Wegscheider}
\end{equation}
Equivalently, the cycle affinity associated with $\mathcal C$ vanishes, $\mathcal A_{\mathcal C} \equiv \ln\frac{k_2k_3k_4k_{-5}}{k_{-2}k_{-3}k_{-4}k_5} = 0 $. Consequently, the internal cycle cannot contribute to the entropy production rate.
We impose Eq. (\ref{eq:Wegscheider}) also for the open CRN with chemostatted species $S$ and $P$.

\remark{Remark (time-averages):}
If the solution of the RRE (\ref{eq:RRE}) for the internal species describes a periodic motion with period $\mathcal T$, the system is said to be in a periodic steady-state. However, if the solution for the internal species does not change in time anymore, it has relaxed to a fixed point, which is denoted as the system has reached a steady-state. In the following, we treat a steady-state like a periodic steady-state with an arbitrary period $\mathcal T$. In a periodic steady-state, we define the time-average of an observable $\mathcal O$ as follows,
\begin{equation}
    \braket{\mathcal O} \equiv \frac{1}{\mathcal T} \int_0^{\mathcal T} \mathcal O(t)dt\,.
\end{equation}
The semi-grand Gibbs free-energy is a state function, hence, has an exact differential. Thus, integration of Eq. (\ref{eq:1stCurly}) over a period $\mathcal T$ leads to
\begin{equation}
    \braket{\dot\Sigma} = \braket{\dot W}\,,\label{eq:EPRisW}
\end{equation}
thus, in the following, we focus on the time-averaged work-rate instead of the entropy production rate.

\section{Numerical Analysis of the Hopf Bifurcation and Thermodynamic Response}
\label{sec:Numerics}

\begin{figure*}[t]
    \centering
    \includegraphics[width=\figwidth]
        {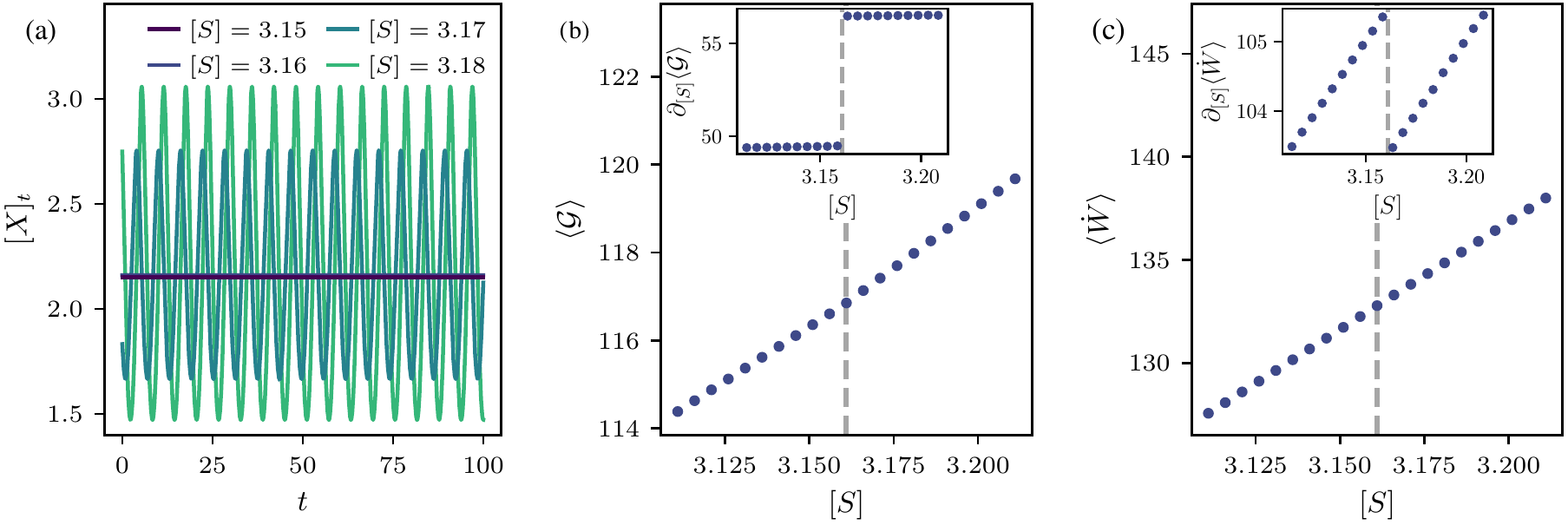}
   \caption{
Hopf bifurcation and thermodynamic signatures in the minimal CRN (\ref{eq:CRN}). 
(a) Time evolution of the concentration $[X]$ for different values of the control parameter $[S]$, with oscillations observed for $[S]>[S_c]\simeq 3.161$. 
(b) Time-averaged semi-grand Gibbs free energy $\braket{\mathcal G}$ as a function of $[S]$. Inset: the derivative $\partial_{[S]}\braket{\mathcal G}$ is discontinuous at the critical value $[S_c]\simeq 3.161$ and increases across the bifurcation. 
(c) Time-averaged non-conservative work rate $\braket{\dot W}$ as a function of $[S]$. Inset: the derivative $\partial_{[S]}\braket{\dot W}$ is discontinuous at $[S_c]$ and decreases across the bifurcation. 
The grey dashed line indicates $[S_c]$. 
Parameters (arbitrary units): $k_1=k_2=k_3=k_4=k_5=1$, $k_{-1}=k_{-2}=k_{-3}=k_{-4}=10^{-2}$, $k_{-5}=10^{-6}$, and $[P]=1$. Concentrations are expressed in arbitrary units.
}    
\label{fig:Osci}
\end{figure*}

We examine the dynamics and thermodynamics of CRN (\ref{eq:CRN}) by numerically solving RRE (\ref{eq:RRE}) in the long-time limit. From these solutions, we evaluate time-averaged thermodynamic observables, in particular the semi-grand Gibbs free energy, Eq. (\ref{eq:SemigrandGibbs}), and the non-conservative work rate, Eq. (\ref{eq:WorkRate}), see Fig.~\ref{fig:Osci}. 
Consistent with typical experimental conditions in continuously driven chemical reactors, we fix the kinetic constants $\{k_\rho\}$ and the chemostat concentration $[P]$, while varying $[S]$ as the control parameter. The parameter values are chosen to illustrate the main findings.

The inspection of numerical simulations performed upon varying $[S]$ shows that the concentration of species $X$ transitions from a steady state to a periodic steady state when crossing a critical value $[S_c] \simeq 3.161$, Fig.~\ref{fig:Osci}a. The same behaviour is observed for species $Y$ and $Z$ (data not shown). This transition from a stable fixed point to sustained oscillations identifies a Hopf bifurcation in the deterministic dynamics.

We next examine how this dynamical transition is reflected in thermodynamic observables. The time-averaged semi-grand Gibbs free energy exhibits a change in behavior at $[S_c]$, see Fig.~\ref{fig:Osci}b. In particular, its derivative with respect to the control parameter shows a discontinuity at the bifurcation (inset of Fig.~\ref{fig:Osci}b), corresponding to an increased sensitivity of $\braket{\mathcal G}$ beyond the onset of oscillations.

A corresponding change is observed for the time-averaged work rate, Fig.~\ref{fig:Osci}c. Here, the derivative $\partial_{[S]}\braket{\dot W}$ displays a discontinuity of opposite sign at $[S_c]$ (inset of Fig.~\ref{fig:Osci}c), indicating a reduced sensitivity of the work rate—and therefore of the dissipation, see Eq.~(\ref{eq:EPRisW})—to variations in $[S]$ in the oscillatory regime.

Taken together, these results show that the Hopf bifurcation is accompanied by a well-defined thermodynamic response in time-averaged observables. In the present parameter regime, this response takes the form of kink-like nonanalyticities, manifested as discontinuities in first derivatives with respect to the control parameter.\cite{nguy18,fala18, herp18, meib24,gopa25}

These observations indicate that the thermodynamic behavior is closely tied to the dynamical transition. In the following sections, we analyze this connection analytically by studying the dynamics in the vicinity of the bifurcation.

\remark{Remark (Hopf bifurcation):} 
We additionally verified the stability of the fixed point by numerically computing the spectrum of the Jacobian matrix of the RRE (\ref{eq:RRE}) and evaluating the Hopf condition, Eq.~(\ref{eq:HopfCondition}), as discussed in Sec.~\ref{sec:NormalFormMapping}. This yields the estimate $[S_c] \simeq 3.161$, confirming the Hopf bifurcation.

\section{Normal-form analysis near the Hopf bifurcation}
\label{sec:TDObservables}

Here, we analyze the Hopf bifurcation of the reaction-rate equations (\ref{eq:RRE}) by means of a normal-form reduction and examine the resulting behavior of the thermodynamic observables $\braket{\mathcal G}$ and $\braket{\dot W}$.

\subsection{Normal form mapping}
\label{sec:NormalFormMapping}

The RRE (\ref{eq:RRE}) defines a three-dimensional dynamical system. Near the bifurcation, their local behavior can be characterized using a mapping to the normal-form for a Hopf bifurcation \cite{craw91}. In the vicinity of the bifurcation, the Jacobian matrix of the vector field evaluated in the fixed point has a real negative eigenvalue, denoted $\Lambda_0< 0$, and a pair of complex conjugated eigenvalues,
\begin{equation}
    \Lambda_\pm \equiv \Lambda_R \pm i \Lambda_I\,.\label{eq:ImaginaryEigenvalue}
\end{equation}
This pair of complex eigenvalues satisfies the conditions of the Hopf bifurcation theorem\cite{guck83,wigg03},
\begin{equation}
    \begin{split}
        \Lambda_R\big |_{\Delta S = 0} &= 0\,,\\
        \frac{d}{d\Delta S}\Lambda_R\big |_{\Delta S = 0} &\neq 0\,,\\
        \Lambda_I\big |_{\Delta S = 0} &\neq 0\,,
    \end{split}\label{eq:HopfCondition}
\end{equation}
where $\Delta S \equiv [S]-[S_c]$ denotes the control parameter. Eq. (\ref{eq:HopfCondition}) characterizes the transition from a stable fixed point to an unstable fixed point with the onset of oscillations, i.e., crossing the real-axis with non-vanishing velocity and non-vanishing imaginary part.
For the present system, these conditions permit a reduction to the cubic Hopf normal form\cite{craw91},
\begin{equation}
    \begin{split}
        \partial_t r_t &= \Re(\mathfrak l)r_t (r_t^2 + \Lambda_R/\Re(\mathfrak l))\,,\\
        \partial_t \theta_t &= \Lambda_I - \Im(\mathfrak l) r_t^2\,,
\end{split}\label{eq:CubicNormalFormPolar}
\end{equation}
where $r_t$ denotes the radial component of the emerging oscillations around the unstable fixed point -- termed limit-cycle -- at time $t$ and $\theta_t$ the angular component, $\Re(\mathfrak l)$ denotes the real-part of $\mathfrak l$ and $\Im(\mathfrak l)$ its imaginary part, respectively. 
The coefficient $\mathfrak l$ originates from projecting the three-dimensional vector-field, the right hand side of RRE (\ref{eq:RRE}), onto the subspace of the pair of complex-conjugated eigenvalues and further technicalities as outlined in Appendix \ref{app:NormalForm}. 
Eq. (\ref{eq:CubicNormalFormPolar}) is termed the normal form for a Hopf bifurcation expressed in polar coordinates. 

From Eq. (\ref{eq:CubicNormalFormPolar}), the limit cycle radius for a supercritical bifurcation, $\Re(\mathfrak l) < 0$, is given by
\begin{equation}
    r_{LC} \equiv \sqrt{- \frac{\Lambda_R}{\Re(\mathfrak l)}} \propto \sqrt{\Delta S}\,,\label{eq:rLC}
\end{equation}
and the angular frequency,
\begin{equation}
\omega_{LC} \equiv \Lambda_I  + \Im (\mathfrak l)\frac{\Lambda_R}{\Re(\mathfrak l)} \propto const. + \Delta S\,,\label{eq:omegaLC}
\end{equation}
where the proportionalities follow from Eq. (\ref{eq:HopfCondition}).

In the normal-form coordinates, the limit-cycle oscillations are parameterized as
\begin{equation}
    \begin{pmatrix}
        U_t\\ V_t
    \end{pmatrix} = r_{LC} \begin{pmatrix}
        \cos(\omega_{LC}t)\\ \sin(\omega_{LC}t)
    \end{pmatrix}\,,\label{eq:LimitCycle}
\end{equation}
with $r_{LC}$ and $\omega_{LC}$ given in Eq. (\ref{eq:rLC}) and (\ref{eq:omegaLC}), respectively.
The normal-form mapping as discussed in Appendix \ref{app:NormalForm} does not introduce any time dependence, neither does the back-transformation. Thus, the time-dependence in the oscillating concentrations is solely described by the parameterized limit-cycle, Eq. (\ref{eq:LimitCycle}),
\begin{equation}
    \begin{pmatrix}
        [X]_t\\
        [Y]_t\\
        [Z]_t
    \end{pmatrix} =   \begin{pmatrix}
        [X_*]\\
        [Y_*]\\
        [Z_*]
    \end{pmatrix} + \begin{pmatrix}
        \delta X_t\\
        \delta Y_t\\
        \delta Z_t
    \end{pmatrix} \,,\label{eq:BackTransformedConcentration}
\end{equation}
where $[\alpha_*]$ with $\alpha \in \mathcal S_i$ denotes the fixed point concentration,
 $\delta \alpha_t \equiv  F_\alpha(U_t,V_t)$ and we summarized all steps of the back-transform in $F_\alpha(U_t,V_t)$, $\alpha \in \mathcal S_i$; see Appendix \ref{app:NormalForm}. 

\subsection{Expansion of observables}
\label{sec:Thermo}

We now analyze the behavior of observables of CRN (\ref{eq:CRN}) in the vicinity of the bifurcation. From Eq. (\ref{eq:BackTransformedConcentration}), it follows that the time-average of any smooth observable $\mathcal O$ evaluated along the limit cycle does not depend explicitly on the limit-cycle frequency,
\begin{equation}
\begin{split}
    \braket{\mathcal O}
    & =  \frac{\omega_{LC}}{2\pi} \int_0^{2\pi/\omega_{LC}} \mathcal O([\boldsymbol \alpha_*]+\boldsymbol{\delta \alpha}_t)dt\\
    & =  \frac{1}{2\pi} \int_0^{2\pi} \mathcal O\big([\boldsymbol \alpha_*]+\boldsymbol F(r_{LC}\cos(\tilde t), r_{LC}\sin(\tilde t))\big)d\tilde t\,,
\end{split}
\end{equation}
where a bold font denotes the vector of the given quantity with components labeled by $\alpha \in \mathcal S_i$, e.g., $[\boldsymbol \alpha]_t \equiv ( [X]_t,\,[Y]_t,\,[Z]_t)$. 

Any time-average $\braket{\mathcal O}$ implicitly depends on the control parameter $\Delta S$ through the limit cycle radius $r_{LC} \propto \sqrt{\Delta S}$. Furthermore, any deviations $\boldsymbol{\delta\alpha}_t$ from the fixed points vanish for $\Delta S = 0$. Writing
\begin{equation}
\begin{split}
\boldsymbol{\delta \alpha}_{t/\omega_{LC}} &=     \boldsymbol F(\underbrace{r_{LC}\cos(t)}_{U_{t/\omega_{LC}}},\underbrace{r_{LC}\sin(t)}_{V_{t/\omega_{LC}}}) \\&= \boldsymbol c^{(1/2)}_t \sqrt{\Delta S} + \boldsymbol c^{(1)}_t \Delta S + \ldots\,,
\end{split}\label{eq:Deviations}
\end{equation}
it follows that
\begin{equation}\small
\begin{split}    
    \mathcal O([\boldsymbol\alpha_*] + \boldsymbol{\delta\alpha}_{t/\omega_{LC}}) - \mathcal O_* &\approx  \sqrt{\Delta S}\, \boldsymbol c^{(1/2)}_t \cdot \nabla \mathcal O_*\big|_{\Delta S = 0}\\
    & \quad+ \Delta S\,\big\{\boldsymbol c^{(1)}_t\cdot \nabla\mathcal O_*\big|_{\Delta S = 0}\\
    & \quad+\frac 12\boldsymbol c^{(1/2)}_t \cdot\nabla^2 \mathcal O_*\big|_{\Delta S = 0}\,\boldsymbol c^{(1/2)}_t \big\}\\ 
    &\quad+ \ldots\,,
\end{split}\label{eq:DiffState}
\end{equation}
where $\mathcal O_* \equiv \mathcal O([\boldsymbol \alpha_*])$, $\nabla \equiv (\ldots\partial_{[\alpha]}\ldots)$ and $\nabla^2 \mathcal O_*$ denotes the Hessian matrix. 

The coefficients $\boldsymbol c_t^{(i)}$ are determined entirely by the dynamics through normal-form transformation $\boldsymbol F(U_t,V_t)$ and are therefore independent of the specific observable under consideration. Since $\boldsymbol F(U_t,V_t)$ is a polynomial without a constant term (Appendix~\ref{app:NormalForm}), the leading coefficient $\boldsymbol c_t^{(1/2)}$ arises from terms linear in $U_t$ and $V_t$. Using the limit-cycle parametrization (\ref{eq:LimitCycle}), it follows that
\begin{equation}
    \int_0^{2\pi}\boldsymbol c_t^{(1/2)}dt=\boldsymbol0\,.
    \label{eq:c12Time}
\end{equation}
Consequently, the leading contribution proportional to $\sqrt{\Delta S}$ in Eq.~(\ref{eq:DiffState}) vanishes upon time averaging. By contrast, $\boldsymbol c_t^{(1)}$ contains quadratic combinations of $U_t$ and $V_t$, whose average generally does not vanish. Averaging Eq.~(\ref{eq:DiffState}) over one period therefore yields
\begin{equation}\small
\begin{split}
    \braket{\mathcal O}-\mathcal O_*
    &\approx
    \Delta S
    \Bigg\{
    \frac{1}{2\pi}
    \int_0^{2\pi}
    \boldsymbol c_t^{(1)}dt
    \cdot
    \nabla\mathcal O_*
    \Big|_{\Delta S=0}
\\
    &\qquad\qquad
    +
    \frac{1}{4\pi}
    \int_0^{2\pi}
    \boldsymbol c_t^{(1/2)}
    \cdot
    \nabla^2\mathcal O_*
    \Big|_{\Delta S=0}
    \boldsymbol c_t^{(1/2)}
    dt
    \Bigg\}
\\
    &\quad+\ldots .
\end{split}
\label{eq:DiffStateTime}
\end{equation}
Thus, the onset of oscillations generically generates a contribution linear in the distance from the bifurcation, $\braket{\mathcal O}-\mathcal O_*\propto\Delta S$. This reproduces the generic scaling structure recently derived for smooth state observables near a supercritical Hopf bifurcation in Ref.~\onlinecite{remlein2026}.

In the next section, we approximate CRN (\ref{eq:CRN}) in an irreversible limit in order to obtain an approximation of the dynamics, RRE (\ref{eq:RRE}). 
In that limit, we obtain analytical expressions for the coefficients $\boldsymbol c_t^{(i)}$ and for the fixed point $[\boldsymbol \alpha_*]$. This enables us to evaluate Eq. (\ref{eq:DiffStateTime}) for $\mathcal G$ and $\dot W$ and to estimate the direction of the jump in the time-averaged thermodynamic observables.

\remark{Remark (analytic fixed point):}
Due to the Hopf conditions discussed in Sec.~\ref{sec:NormalFormMapping}, the Jacobian of the vector field with respect to the variables $[\boldsymbol \alpha]$ has no zero eigenvalue and is therefore invertible at the bifurcation. By the implicit function theorem, the fixed point $[\boldsymbol \alpha_*]$ of mass-action dynamics depends analytically on the control parameter $\Delta S$ in the vicinity of the bifurcation. 
Consequently, for the observables $\mathcal G$ and $\dot W$ that depend analytically on the concentrations $[\boldsymbol \alpha]$, their fixed-point contribution $\mathcal G_*$ and $\dot W_*$ inherits this analyticity in $\Delta S$. Any nonanalyticity must therefore arise from the onset of oscillations, see Eq.~(\ref{eq:DiffStateTime}).

\section{Analytical evaluation via an irreversible approximation}
\label{sec:Irreversible}

We here non-dimensionalize RRE (\ref{eq:RRE}) in order to identify the independent system parameters and introduce an irreversible approximation. We start with the following rescaling,
\begin{equation}
    [\alpha] = \lambda_\alpha [\tilde \alpha]\,,\quad t = \lambda_t \tilde t\,,\label{eq:Nondimension}
\end{equation}
where $[\tilde \alpha]$ denotes the dimensionless concentration of species $\alpha \in \mathcal S$, and $\tilde t$ the dimensionless time, respectively. Dimensionless quantities are retrieved by dividing each physical quantity by its corresponding characteristic reference value. This is a standard approach in CRN analysis, which allows gaining generality.

Requiring that the rescaled RRE should still represent a reversible CRN and that the new kinetic constants $\{\kappa_\rho\}$ should be as close to the numerical regime examined in Sec. \ref{sec:Numerics} as possible, we find that $\kappa_1 = \kappa_2 = \kappa_{-4} = \kappa_5 = 1$ and only $\kappa_{-1}$, $\kappa_{-2}$, $\kappa_{3}$, $\kappa_{-3}$, $\kappa_{4}$, and $\kappa_{-5}$ are independent variables; see Appendix \ref{app:Irreversible} for the derivation and definition of the $\{\kappa_\rho\}$.

Motivated by the numerical values of the $\{\kappa_\rho\}$ consistent with the regime examined in Sec.~\ref{sec:Numerics}, see Eq.~(\ref{eq:NewConstants}), we consider the following approximation,
\begin{equation}
    \kappa_{-1},\kappa_{-2},\kappa_{-3}, \kappa_{-5} \ll 1\,.
\label{eq:IrreversibleLimit}
\end{equation}
This limit results in a simplified irreversible CRN that preserves the leading-order dynamical structure of the full reversible CRN. The controlled approach to this limit is supported numerically in Appendix~\ref{app:irr_convergence}, where we show that the critical Hopf value and the thermodynamic response of the reversible system converge toward the irreversible-limit predictions as the backward reaction rates are reduced. The corresponding CRN reads
\begin{equation}
\begin{split}
S+X &\ch{->[\phantom{$\delta$}]} 2X\,,\\
X   &\ch{->[\phantom{$\delta$}]} Z\,,\\
Z   &\ch{->[$\delta$]} Y\,,\\
Y   &\crn{}{\eta} P\,,\\
X+Y &\ch{->[\phantom{$\delta$}]} P+Y\,,
\end{split}
\label{eq:CRNirr}
\end{equation}
where unlabeled arrows correspond to a kinetic constant equal to unity. We further introduced $\delta \equiv \kappa_{3}$ and $\eta \equiv \kappa_{4}$ for convenience. Note that the rescaled kinetic constants are positive, i.e., $\delta,\eta>0$.

\subsection{Hopf bifurcation of the irreversible CRN}
\label{sec:HopfIrr}

The RRE of the $\mathcal S_i$ species corresponding to the irreversible CRN (\ref{eq:CRNirr}) reads
\begin{equation}
    \frac{d}{dt}\begin{pmatrix}
[X]_t\\
[Y]_t\\
[Z]_t\end{pmatrix} = \begin{pmatrix}
[X]_t([S] - 1 - [Y]_t)\\
\delta [Z]_t - \eta [Y]_t + [P]\\
[X]_t -  \delta[Z]_t
\end{pmatrix} \,,\label{eq:RREirr}
\end{equation}
where we drop the $\tilde {}$-notation indicating rescaled quantities, see Eq. (\ref{eq:Nondimension}). 
We assume $\delta$ and $\eta$ as fixed parameters, while $[S]$ and $[P]$ serve as control parameters. 
In Appendix \ref{app:LinearStability}, we perform a comprehensive linear stability analysis of Eq. (\ref{eq:RREirr}). Here, we discuss the main aspects regarding the Hopf bifurcation.

At a Hopf bifurcation, the eigenvalues of the linear stability analysis are characterized by the following characteristic polynomial,
\begin{equation}
    \chi_J(\Lambda) = - (\Lambda - \Lambda_0)(\Lambda^2 + \Lambda_I^2) = -\Lambda^3 + \Lambda_0 \Lambda^2 - \Lambda_I^2 \Lambda + \Lambda_0 \Lambda_I^2\,,\label{eq:CharPolyHopf}
\end{equation}
where $\Lambda_0$ denotes the eigenvalue corresponding to the stable direction of the fixed point, and $\Lambda_I$ the imaginary part of the purely complex conjugate pair of eigenvalues, see Eq. (\ref{eq:HopfCondition}). 
In the parameter regime, where the Hopf bifurcation could appear, the characteristic polynomial corresponding to RRE (\ref{eq:RREirr}) reads
\begin{equation}
    \chi_J(\Lambda)= -\Lambda^3 - (\eta+\delta)\Lambda^2 - \eta \delta \Lambda - \delta \eta ([S]-1) + \delta [P]\,,\label{eq:CharPoly}
\end{equation}
see Appendix \ref{app:Eigenvalues} for the associated Jacobian matrix.
Equating the coefficients in Eq. (\ref{eq:CharPolyHopf}) and (\ref{eq:CharPoly}) yields
\begin{subequations}
\begin{align}
    \Lambda_0 &= - (\eta + \delta)\,, \\
    \Lambda_I &= \sqrt{\eta \delta}\,,\\
    [S_\mathrm{HB}] &\equiv 1 + \frac{[P]}\eta + \eta+ \delta \,.\label{eq:HopfLine}
\end{align}    
\end{subequations}
Eq.~(\ref{eq:HopfLine}) describes the Hopf line in the $[S]-[P]$ plane. For convenience we suppress the $[P]$-dependence of $[S_\mathrm{HB}]$. We refer to this parameterized line as Hopf-line.
Furthermore, we neglect a negative solution for $\Lambda_I$ since it leads to the same complex eigenvalues, $\Lambda_\pm = \Lambda_R\pm i \Lambda_I$ (see Eq. (\ref{eq:ImaginaryEigenvalue})), where, for $[S] = [S_\mathrm{HB}]$, $\Lambda_R = 0$.

Using Mathematica\cite{Mathematica} to solve Eq. (\ref{eq:CharPoly}) for the general eigenvalues, Eq. (\ref{eq:Eigenvalues}), and performing a Taylor-expansion for $\Delta S \equiv [S]-[S_\mathrm{HB}] \ll 1$ gives
\begin{equation}
    \begin{split}
        \Lambda_0 &= -(\delta + \eta) - \frac{\eta \delta}{(\eta + \delta)^2 + \eta\delta}\Delta S\,,\\
        \Lambda_R &= \frac{\eta\delta}{2((\eta + \delta)^2 + \eta\delta)}\Delta S \,,\\
        \Lambda_I &= \sqrt{\eta\delta} +\frac{\sqrt{\eta\delta}(\eta + \delta)}{2((\eta + \delta)^2 + \eta\delta)}\Delta S \,.
    \end{split}\label{eq:EigenvaluesHopf}
\end{equation} 
These eigenvalues satisfy the conditions formulated in Eq. (\ref{eq:HopfCondition}). Thus, the irreversible CRN (\ref{eq:CRNirr}) undergoes a Hopf bifurcation.
Furthermore, Eq. (\ref{eq:ImaginaryEigenvalue}) together with Eq. (\ref{eq:EigenvaluesHopf}) determines the eigenvalues $\Lambda_0$, $\Lambda_+$ and $\Lambda_-$ of the linear stability analysis in the vicinity of the Hopf bifurcation.

\subsection{Evaluation of $C_\mathcal{G}$ and $C_{\dot{W}}$}

Here, we perform the normal form mapping for the irreversible CRN (\ref{eq:CRNirr}) and derive approximations for the dynamical quantities discussed in Sec. \ref{sec:Thermo}. To do so, we use the analytical findings for the Hopf bifurcation of the irreversible CRN (\ref{eq:CRNirr}) presented in Sec. \ref{sec:HopfIrr} and Appendix \ref{app:LinearStability}, as well as Mathematica to perform the normal form mapping regarding the irreversible RRE (\ref{eq:RREirr}) as discussed in Sec. \ref{sec:NormalFormMapping} and outlined in Appendix \ref{app:NormalForm}. The supporting results obtained with Mathematica are shown in Appendix \ref{app:MathematicaResults}.

The dynamical expansion coefficients $\boldsymbol c_t^{(i)}$ of the irreversible RRE (\ref{eq:RREirr}) satisfy
\begin{equation}
    \int_0^{2\pi} \boldsymbol c_t^{(1/2)}dt = 0\,,~\text{and}~\int_0^{2\pi} \boldsymbol c_t^{(1)}dt = \boldsymbol 0\,, 
\end{equation}
compare Eq. (\ref{eq:App:c12}) and (\ref{eq:App:c1}). Furthermore, the Hessian matrices of $\mathcal G$ and $\dot W$ are diagonal matrices. Taking everything into account and performing the time-average, Eq. (\ref{eq:DiffStateTime}) leads to
\begin{equation}\small
\begin{split}    
    \braket{\mathcal O} - \mathcal O_*& \equiv \frac{1}{2\pi}\int_0^{2\pi}\{\mathcal O([\boldsymbol\alpha_*] + \boldsymbol{\delta\alpha}_{t/\omega_{LC}}) - \mathcal O([\boldsymbol\alpha_*])\}dt\\
    & \simeq\Delta S\frac 1{4\pi} \sum\limits_{\alpha \in \mathcal S_i}\big( \int_0^{2\pi}(c^{(1/2)}_t{}_\alpha)^2dt\big)\partial^2_{[\alpha]} \mathcal O([\boldsymbol\alpha_*])\,.
\end{split}\label{eq:DiffStateTimeIrr}
\end{equation}

The second derivative of the semi-grand Gibbs free energy $\mathcal G$, Eq. (\ref{eq:SemigrandGibbs}), reads
\begin{equation}
    \partial^2_{[\alpha]}\mathcal G = \frac 1{[\alpha_*]} > 0\,,
\end{equation}
for $\alpha \in \mathcal S_i$. Thus, the behavior in the vicinity of the bifurcation is given by
\begin{equation}
    \braket{\mathcal G} - \mathcal G_* \approx \begin{cases}
        C_{\mathcal G}\Delta S &,\Delta S \geq 0 \\ 0 &, \Delta S < 0
    \end{cases}\,,\label{eq:Gdiff}
\end{equation}
with $C_{\mathcal G} \equiv \sum_{\alpha \in \mathcal S_i}\big( \int_0^{2\pi}(c^{(1/2)}_t{}_\alpha)^2dt\big)/4\pi [\alpha_*] > 0$, where $\int_0^{2\pi}(c^{(1/2)}_t{}_\alpha)^2dt > 0$ has been used. The exact expression for $C_\mathcal{G}$ is given in Eq. (\ref{eq:App:CG}). 
Thus, Eq.~(\ref{eq:Gdiff}) shows that the derivative of the semi-grand Gibbs free energy exhibits an upward jump at the Hopf bifurcation. Since $C_{\mathcal G}>0$, the sign of this jump is fixed by the network structure and remains positive throughout the regime described by the irreversible approximation.

The work rate depends only on the concentration of the $X$-species,
\begin{equation}
\dot W([X]) = ([S] [X] - \kappa_{-1} [X]^2)(\ln [S] - \ln [P] + \mu_S^\circ - \mu_P^\circ)\,.
\end{equation}
The second derivative reads
\begin{equation}
\begin{split}
    \partial^2_{[X]}\dot W([X_*]) &= - 2\kappa_{-1} (\ln [S_{\mathrm{HB}}] - \ln [P] + \mu_S^\circ - \mu_P^\circ) \\
    &=  - 2\kappa_{-1} \big(\ln\frac{\eta + \eta\delta + \eta^2 + [P]}{[P]}- \ln \eta\kappa_{-5} \kappa_{-1}\big)\\
    & < 2\kappa_{-1} \ln \eta\kappa_{-5} \kappa_{-1}\,,\label{eq:WorkRateSecondDerivative}
\end{split}
\end{equation}
where the standard chemical potentials have been expressed in terms of the kinetic constants via Eq.~(\ref{eq:LDB}).
Furthermore, consistency with the irreversible approximation requires
\begin{equation}
\eta < \frac{1}{\kappa_{-1}\kappa_{-5}}\,.\label{eq:RateCondition}
\end{equation}
A violation of Eq. (\ref{eq:RateCondition}) would imply a diverging kinetic constant $\eta$ in the irreversible approximation, which leads to nonphysical phenomena as pointed out in Appendix \ref{app:WorkRate}.
Plugging Eq. (\ref{eq:RateCondition}) into Eq. (\ref{eq:WorkRateSecondDerivative}) yields
\begin{equation}
    \partial^2_{[X]}\dot W([X_*]) < 0\,.
\end{equation}
Thus, when crossing the bifurcation, the work rate behaves as
\begin{equation}
    \braket{\dot W} - \dot W_* \approx \begin{cases}
        -C_{\dot W}\Delta S &, \Delta S\geq 0\\ 0&, \Delta S <0
    \end{cases}\,,\label{eq:Wdiff}
\end{equation}
with $C_{\dot W} \equiv \kappa_{-1} (\ln [S_{\mathrm{HB}}] - \ln [P] + \mu_S^\circ - \mu_P^\circ)\int_0^{2\pi}(c^{(1/2)}_t{}_X)^2dt/2\pi > 0$, where $\int_0^{2\pi}(c^{(1/2)}_t{}_X)^2dt > 0$ has been used.
The exact expression of $C_{\dot W}$ is given in Eq. (\ref{eq:App:CW}). 

Thus, Eq.~(\ref{eq:Wdiff}) shows that the derivative of the work rate exhibits a downward jump at the Hopf bifurcation. Within the irreversible approximation, this behavior follows generically from the condition $\partial^2_{[X]}\dot W([X_*])<0$ and therefore does not depend on the specific parameter values considered in Sec.~\ref{sec:Numerics}.

\subsection{Summary and comparison of thermodynamic behavior}
\label{sec:IrreversibleSummary}

Sec.~\ref{sec:TDObservables} identified the mechanism underlying the nonanalytic behavior of the thermodynamic observables in the present model, while the irreversible approximation of the minimal CRN (\ref{eq:CRN}) developed here captures its leading qualitative features and permits explicit analytical estimates.
To this end, we introduced an irreversible CRN (\ref{eq:CRNirr}) as a leading-order approximation to the dynamics of the reversible CRN (\ref{eq:CRN}). This limit corresponds to neglecting the reactions listed in Eq.~(\ref{eq:IrreversibleLimit}), enabling analytical predictions for the phase-space structure (Appendix~\ref{app:LinearStability}) and the resulting dynamics governed by the irreversible RRE (\ref{eq:RREirr}).

The irreversible CRN (\ref{eq:CRNirr}) exhibits a Hopf bifurcation. The corresponding Hopf line in phase space, as well as the behavior of the eigenvalues in its vicinity, can be determined analytically. Building on these dynamical results within the irreversible approximation, we estimated the dominant contributions to the thermodynamic observables associated with the onset of oscillations.

\begin{figure}[h!]
    \centering
    \includegraphics[width=\columnwidth]
        {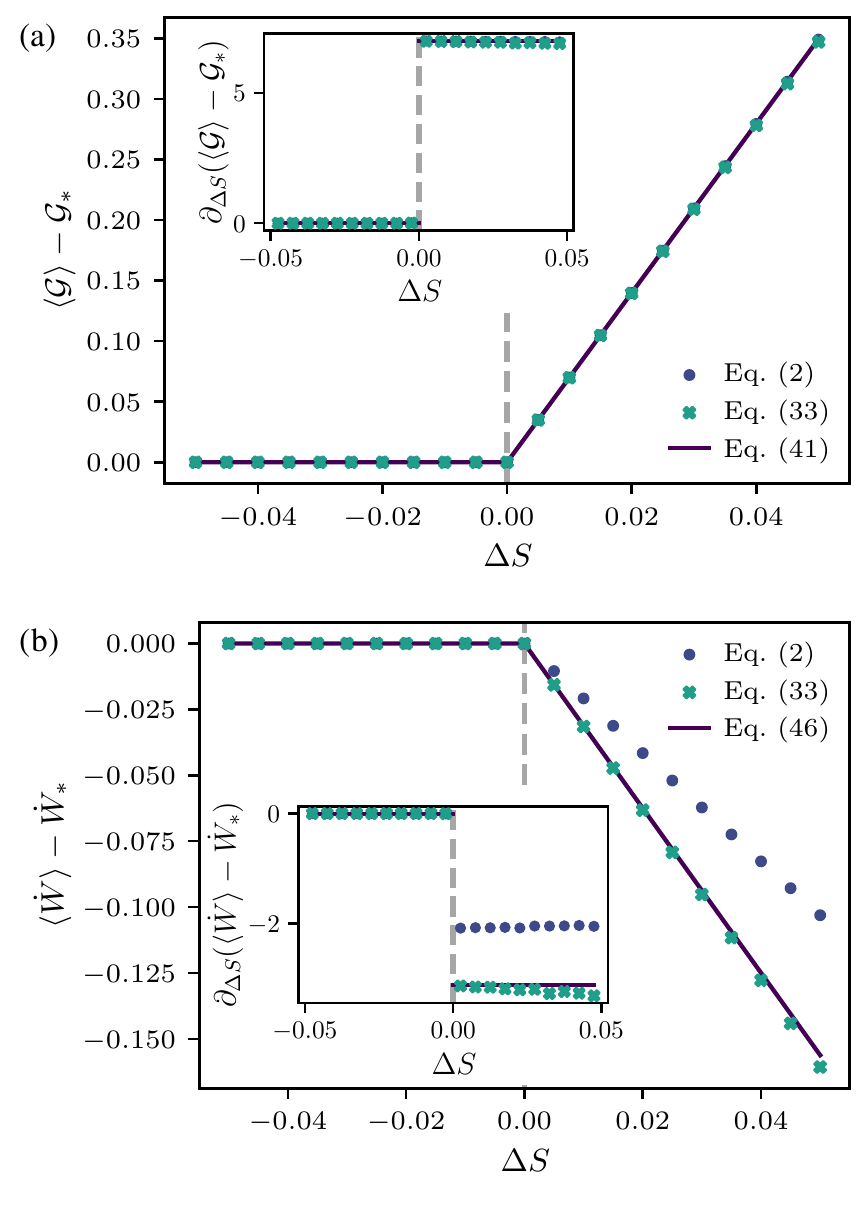}     
    \caption{
    Comparison of the reversible CRN (\ref{eq:CRN}), the irreversible CRN (\ref{eq:CRNirr}), and the analytical solution in the vicinity of the Hopf bifurcation. 
    Here, $\Delta S \equiv [S]-[S_c]$ for the reversible RRE~(\ref{eq:RRE}), and $\Delta S \equiv [S]-[S_\mathrm{HB}]$ for the irreversible RRE~(\ref{eq:RREirr}). The analytical solutions are obtained using Eq.~(\ref{eq:Gdiff}) and Eq.~(\ref{eq:Wdiff}).  
    (a) Time-averaged semi-grand Gibbs free energy $\braket{\mathcal G} - \mathcal G_*$ as a function of the control parameter $\Delta S$. 
    Inset: the derivative $\partial_{\Delta S}(\braket{\mathcal G} - \mathcal G_*)$ is discontinuous at the bifurcation.
    (b) Time-averaged non-conservative work rate $\braket{\dot W} - \dot W_*$ as a function of $\Delta S$. 
    Inset: $\partial_{\Delta S}(\braket{\dot W} - \dot W_*)$ is discontinuous at the bifurcation.
    Parameters (arbitrary units): $\kappa_1 = \kappa_2 = \kappa_3 = \kappa_4 = \kappa_5 = 1$, $\kappa_{-1} = \kappa_{-2} = \kappa_{-3} = \kappa_{-4} = 10^{-2}$, $\kappa_{-5} = 10^{-4}$, and $[\tilde P] = 10^{-2}$. $\Delta S$ is expressed in arbitrary units.
    }
    \label{fig:Irr}
\end{figure}

The general structure of the $[S]-[P]$ phase space of the reversible and irreversible CRNs, Eq.~(\ref{eq:CRN}) and (\ref{eq:CRNirr}), respectively, is in qualitative agreement, as shown in Appendix~\ref{app:PhaseSpace}. Furthermore, the theoretical predictions for the semi-grand Gibbs free energy, Eq.~(\ref{eq:Gdiff}), obtained within the irreversible approximation, agree well with numerical data from both the irreversible and reversible RRE, as shown in Fig.~\ref{fig:Irr}a. In contrast, the theoretical prediction for the work rate, Eq.~(\ref{eq:Wdiff}), agrees with the irreversible RRE but deviates from the reversible RRE, Fig.~\ref{fig:Irr}b.

This discrepancy can be traced back to the numerical observation that, for solutions of the reversible RRE (\ref{eq:RRE}), it holds that
\begin{equation}
    \int_0^{2\pi} \boldsymbol c_t^{(1)} dt \neq 0\,. 
    \label{eq:c1Discrepancy}
\end{equation}
Consequently, the first-derivative terms in Eq.~(\ref{eq:DiffStateTime}) contribute to the linear dependence on the control parameter. Although this affects both observables $\mathcal G$ and $\dot W$, we find numerically that the contribution to $\mathcal G$ is negligible compared with the Hessian term involving $\nabla^2 \mathcal G_*$. By contrast, it accounts for the discrepancy in $\braket{\dot W}-\dot W_*$ shown in Fig.~\ref{fig:Irr}b. Importantly, the coefficients $\int_0^{2\pi} c^{(1/2)}_{\alpha;t} dt$ and $\int_0^{2\pi}\big(c^{(1/2)}_{\alpha;t}\big)^2 dt$, which determine the dominant contribution, are quantitatively similar for both the irreversible and the reversible RRE.

To further analyze the irreversible approximation, Appendix~\ref{app:irr_convergence} examines a one-parameter family approaching the limit $\kappa_{-1},\kappa_{-2},\kappa_{-3},\kappa_{-5}\to0$ while preserving the Wegscheider constraint, Eq.(\ref{eq:Wegscheider}). We numerically find that the critical Hopf value $[S_c]$ and the jump coefficients $C_{\mathcal G}$ and $C_{\dot W}$ converge toward the analytical predictions of the irreversible model throughout the explored parameter regime.

In summary, the analytical approach based on the normal-form mapping combined with the irreversible approximation reproduces the two main qualitative features of the thermodynamic observables identified in Sec.~\ref{sec:Numerics}: $i)$ the kink-like dependence and $ii)$ the direction of the kink with respect to the control parameter.
Furthermore, within the irreversible approximation, the direction of the thermodynamic response follows from general properties of the observables. The upward jump of the semi-grand Gibbs free energy is a consequence of the local convexity of $\mathcal G$ around the fixed point, which implies $C_{\mathcal G}>0$, whereas the downward jump of the work rate follows from $\partial_{[X]}^2\dot W([X_*])<0$. These predictions therefore do not rely on the specific parameter values considered in Sec.~\ref{sec:Numerics}.
While this framework captures the leading-order structure of the dynamics of the reversible CRN, additional contributions -- most notably those associated with $\int_0^{2\pi}\boldsymbol c^{(1)}_t dt$ -- lead to quantitative deviations. These corrections are particularly relevant for the work rate, while their effect on the semi-grand Gibbs free energy remains subdominant.

\remark{Remark (numerical evaluation of the time-average of Eq.~(\ref{eq:DiffState})):}
From Eq.~(\ref{eq:Deviations}) and (\ref{eq:c12Time}), it follows that
\begin{equation}
    \braket{\delta \alpha}
    =
    \braket{c_\alpha^{(1)}}\Delta S
    + \ldots\,,
\end{equation}
and
\begin{equation}
    \braket{\delta \alpha^2}
    \simeq
    \braket{\big(c^{(1/2)}_\alpha\big)^2}\Delta S\,,
\end{equation}
for $\alpha \in \mathcal S_i$. Hence, the coefficients $\braket{c^{(1)}_\alpha}$ and $\braket{(c^{(1/2)}_\alpha)^2}$ can be estimated numerically by evaluating $\braket{\delta \alpha}$ and $\braket{\delta \alpha^2}$ along solutions of RRE (\ref{eq:RRE}) or (\ref{eq:RREirr}) and extracting their slopes with respect to $\Delta S$. The discrepancy in the coefficients $\braket{\boldsymbol c^{(1)}}$, see Eq.~(\ref{eq:c1Discrepancy}), propagates through Eq.~(\ref{eq:DiffState}) and gives rise to the observed deviation in $\braket{\dot W}-\dot W_*$.
For the reversible CRN, $\braket{\delta \alpha}$ is numerically found to be linear in $\Delta S$, thus confirming $\braket{\boldsymbol c^{(1/2)}}=0$. Furthermore, the coefficients $\braket{\boldsymbol c^{(1)}}$ are found to be of order $10^{-2}$, whereas they are suppressed to order $10^{-4}$ in the irreversible approximation. This difference reflects a systematic deviation introduced by the irreversible limit. The coefficients $\braket{(c^{(1/2)}_\alpha)^2}$ also show small quantitative differences between the reversible and irreversible CRNs, but remain of the same order of magnitude, taking values in the range $4$--$16$.

\section{Conclusion}
\label{sec:Conclusion}

We have introduced a minimal oscillating chemical reaction network that combines chemical plausibility, thermodynamic consistency, and analytical tractability in a compact setting. The model consists of three internal species coupled to two chemostatted species and involves only uni- and bimolecular reactions while satisfying local detailed balance. This network (Fig.~\ref{fig:CRN}) represents a simplification of the CRN previously identified as the smallest CRN featuring a Hopf bifurcation,\cite{wilh97} and we do not exclude that further simplification might be possible\cite{Banaji_2023}. The present construction provides a chemically explicit realization of an autonomous oscillator that admits a transparent description of key thermodynamic features, namely the semi-grand potential, work, and dissipation.

A central feature of this system is that its behavior near the bifurcation can be analyzed analytically. By combining a normal-form reduction near the Hopf bifurcation with a controlled irreversible approximation, we obtain explicit access to the phase-space structure and the leading contributions governing thermodynamic observables. The irreversible limit preserves the essential dynamical features of the full reversible network while rendering the analysis tractable, thereby providing a controlled asymptotic description in regimes of weak backward reactions.

The resulting framework provides a direct connection between bifurcation structure and thermodynamic response in a chemically explicit model. As an application, we have shown that the onset of oscillations is accompanied by kink-like discontinuities in time-averaged observables, which so far were observed in studies which employed numerical approaches to investigate more elaborated CRN.\cite{fala18,nguy18, herp18, meib24,gopa25} In the present setting, these features can be traced back to the underlying dynamics using the normal-form reduction together with the irreversible approximation.

Beyond this specific example, the model provides a starting point for studying the interplay between oscillatory dynamics, thermodynamic consistency, and energetic costs in driven chemical reaction networks. Its compact structure and analytical accessibility make it a natural basis for extensions to stochastic descriptions, for instance within the chemical master equation, where the same thermodynamically consistent reaction structure and controlled irreversible limit can be used to investigate fluctuation effects and finite-size rounding near dynamical transitions.
More generally, the present work suggests that controlled irreversible approximations can provide a systematic route from thermodynamically consistent reaction networks to analytically tractable descriptions of oscillatory nonequilibrium dynamics. Similar asymptotic reductions may provide further analytical insight into the relation between dynamical transitions and thermodynamic response in more complex networks. In this way, the framework developed here complements ongoing efforts to understand the thermodynamic efficiency and energetic costs of autonomous chemical function in driven reaction networks.\cite{Penocchio2019,amano2022info,pump2_22,elettro23,binks2023,olivieri2024,al-shehimy2024,wu2024,Liang2025,Imines25,Roberts2025,Vesicles26}

\section*{Acknowledgments}
BR, GR, and ME are supported by Project
INTER/ANR/25/19593353-NERD, funded by Fond National de la Recherche (FNR) Luxembourg and Agence Nationale de la Recherche (ANR) France. This work was supported by the European Research Council (ERC-2021-StG 101041933 – KI-NET to G.R.) and the French National program "investments for the future” IdEx Unistra (ANR-10-IDEX-0002), via the Interdisciplinary Thematic Institute ITI-CSC. EP gratefully acknowledges funding from the European Union’s Marie Skłodowska-Curie Actions (MSCA) Postdoctoral Fellowships program under Grant Agreement No. 101268485.

\bibliography{refs.bib}

\appendix

\section{Normal form mapping of 3D vector field}
\label{app:NormalForm}

We here outline the normal form mapping for a 3D vector field, Eq. (\ref{eq:CubicNormalFormPolar}), following Sec. IX.A of Ref. \onlinecite{craw91}. For more technical background we refer to the original publication. Where appropriate, we illustrate the general steps with the irreversible RRE (\ref{eq:RREirr}).

\subsection{Set-up}
We consider a dynamical system characterized by an ordinary differential equation (ODE) that depends on an order parameter $\Delta S$, like the irreversible RRE (\ref{eq:RREirr}) in the vicinity of the bifurcation. Shifting the coordinate system into the stable fixed point,
\begin{equation}
    \boldsymbol{\delta\alpha}_t = [\boldsymbol \alpha]_t - [\boldsymbol \alpha_*]\,,
\end{equation}
the ODE reads,
\begin{equation}
    \frac{d}{dt}\begin{pmatrix}
\delta X_t\\
\delta Y_t\\
\delta Z_t\end{pmatrix} = \underbrace{\begin{pmatrix}
0& -\eta(\delta +  \eta + \Delta S) & 0\\
0 & - \eta & \delta\\
1 & 0 & -  \delta
\end{pmatrix}}_{\equiv L}\begin{pmatrix}
\delta X_t\\
\delta Y_t\\
\delta Z_t\end{pmatrix}+\underbrace{\begin{pmatrix}
-\delta X_t \delta Y_t\\
0\\
0
\end{pmatrix}}_{\equiv \boldsymbol{NL}(\boldsymbol{\delta \alpha_t})} \,,\label{eq:App:RREirr}
\end{equation}
where we defined the matrix $L$ representing the linear part of the vector field, and its nonlinear part $\boldsymbol{NL}(\boldsymbol{\delta \alpha_t})$.

\subsection{Linear terms}
We proceed with a change of basis into the eigenbasis of the matrix $L$. The matrix representing this transformation is denoted $S$ and satisfies,
\begin{equation}
    S^{-1}LS = \begin{pmatrix}
        \Lambda_R & \Lambda_I & 0 \\
        - \Lambda_I& \Lambda_R & 0\\
        0 & 0 & \Lambda_0
    \end{pmatrix}\,,
\end{equation}
where the eigenvalues are given in Eq. (\ref{eq:EigenvaluesHopf}) and the corresponding eigenvectors make up the columns of $S$. In the new coordinates,
\begin{equation}
    \boldsymbol{\delta \alpha^\prime}_t \equiv S^{-1}\boldsymbol{\delta \alpha}_t\,,
\end{equation}
Eq. (\ref{eq:App:RREirr}) takes the following form,
\begin{equation}
    \frac{d}{dt}\begin{pmatrix}
\delta X^\prime_t\\
\delta Y^\prime_t\\
\delta Z^\prime_t\end{pmatrix} =\begin{pmatrix}
        \Lambda_R & \Lambda_I & 0 \\
        - \Lambda_I& \Lambda_R & 0\\
        0 & 0 & \Lambda_0
    \end{pmatrix}\begin{pmatrix}
\delta X^\prime_t\\
\delta Y^\prime_t\\
\delta Z^\prime_t\end{pmatrix}+\underbrace{\begin{pmatrix}
R_1(\delta X^\prime_t,\delta Y^\prime_t,\delta Z^\prime_t)\\
R_2(\delta X^\prime_t,\delta Y^\prime_t,\delta Z^\prime_t)\\
R_3(\delta X^\prime_t,\delta Y^\prime_t,\delta Z^\prime_t)\end{pmatrix}}_{\equiv \boldsymbol R(\boldsymbol{\delta \alpha^\prime}_t)}\,,\label{eq:App:RREirr2}
\end{equation}
where $\boldsymbol R(\boldsymbol{\delta \alpha^\prime_t}) \equiv S^{-1}\boldsymbol{NL}(S\boldsymbol{\delta \alpha^\prime_t})$.

The component functions $R_i(\boldsymbol{\delta \alpha^\prime}_t)$ are  homogeneous, quadratic polynomials. Following the notation of Ref. \onlinecite{craw91}, we define the quadratic coefficients as follows,
\begin{equation}\small
\begin{split}
    R_i(\boldsymbol{\delta \alpha^\prime_t}) &\equiv R_{i,1} \delta X^\prime_t{}^2+R_{i,2} \delta Y^\prime_t{}^2+R_{i,3} \delta X^\prime_t\delta Y^\prime_t\\ & \quad +R_{i,4} \delta X^\prime_t\delta Z^\prime_t+R_{i,5} \delta Y^\prime_t\delta Z^\prime_t+R_{i,6} \delta Z^\prime_t{}^2\,,
\end{split}\label{eq:App:RCoefficients}
\end{equation}
with $i \in \{1,2,3 \}$.

\subsection{Center manifold reduction}
For small displacements, $\delta \alpha^\prime_t\ll1$, $\alpha \in \mathcal S_i$, the decaying degree of freedom can be expressed as a function of the oscillating degrees of freedom, i.e., $\delta Z^\prime_t = h(\delta X^\prime_t,\delta Y^\prime_t)$. To leading order, $h(\delta X^\prime_t,\delta Y^\prime_t)$ is a quadratic polynomial,
\begin{equation}
    h(\delta X^\prime_t,\delta Y^\prime_t) \equiv h_1 \delta X^\prime_t{}^2 + h_2\delta Y^\prime_t{}^2+h_3\delta X^\prime_t\delta Y^\prime_t\,,
\end{equation}\label{eq:App:hPolynomial}
 with coefficients
\begin{equation}
    \begin{split}
        h_3 & = \frac{2 \Lambda_I(R_{3,2} - R_{3,1})+(2 \Lambda_R - \Lambda_0)R_{3,3}}{(2 \Lambda_I)^2 + (2 \Lambda_R - \Lambda_0)^2}\,,\\
        h_1 & = \frac{\Lambda_I h_3 + R_{3,1}}{2\Lambda_R - \Lambda_0} \,,\\
        h_2 & = \frac{- \Lambda_I h_3 + R_{3,2}}{2 \Lambda_R - \Lambda_0}\,.
    \end{split}\label{eq:App:hCoefficients}
\end{equation}
Plugging Eq. (\ref{eq:App:hPolynomial}) into the 3D vector field, Eq. (\ref{eq:App:RREirr2}), eliminates one dimension and leads to a 2D ODE,
\begin{equation}
\begin{split}
    \frac{d}{dt}\begin{pmatrix}
\delta X^\prime_t\\
\delta Y^\prime_t\end{pmatrix} &=\begin{pmatrix}
        \Lambda_R & \Lambda_I  \\
        - \Lambda_I& \Lambda_R 
    \end{pmatrix}\begin{pmatrix}
\delta X^\prime_t\\
\delta Y^\prime_t\end{pmatrix}\\
& \quad+\begin{pmatrix}
R_1(\delta X^\prime_t,\delta Y^\prime_t,\delta Z^\prime_t)\\
R_2(\delta X^\prime_t,\delta Y^\prime_t,\delta Z^\prime_t)\end{pmatrix}\bigg|_{\delta Z^\prime_t = h(\delta X^\prime_t,\delta Y^\prime_t)}\,.
\end{split}\label{eq:App:RREirr3}
\end{equation}

\subsection{Near-identity transformation}
The final steps of the normal form mapping include near-identity transformations,
\begin{equation}
    \boldsymbol \phi_k(\delta X_t^\prime,\delta Y_t^\prime) \equiv \begin{pmatrix}
        \delta X_t^\prime\\\delta Y_t^\prime
    \end{pmatrix} +  \boldsymbol \phi^{(k)}(\delta X_t^\prime,\delta Y_t^\prime)\,,\label{eq:App:NearIdentityTransform}
\end{equation}
where $\boldsymbol \phi^{(k)}(\delta X_t^\prime,\delta Y_t^\prime)$ denotes a vector with component functions that are homogeneous polynomials of order $k$. The first step involves a near-identity transformation of order $2$, which eliminates the quadratic terms in Eq. (\ref{eq:App:RREirr3}). In the appropriate basis (see Ref. \onlinecite{craw91}), the corresponding quadratic $\boldsymbol \phi^{(2)}(\delta X_t^\prime,\delta Y_t^\prime)$ polynomial is characterized by the following coefficients,
\begin{equation}
    \phi_\pm^{(2,l)} \equiv \frac{R_\pm^{(2,l)}}{\Lambda_\pm^{(2,l)}}\,,
\end{equation}
with\cite{craw91}
\begin{equation}
    \begin{split}
        R_+^{(2,0)} &\equiv \frac 14 (R_{1,1}-R_{1,2}-R_{2,3}) + \frac i4 (R_{1,3}+R_{2,1}-R_{2,2}) \equiv \overline{R_-^{(2,2)}}\,,\\
        R_+^{(2,1)} &\equiv \frac 12 (R_{1,1}+R_{1,2}) + \frac i2 (R_{2,1}+R_{2,2}) \equiv \overline{R_-^{(2,1)}}\,,\\
        R_+^{(2,2)} &\equiv \frac 14 (R_{1,1}-R_{1,2}+R_{2,3}) + \frac i4 (R_{2,1}-R_{1,3}-R_{2,2}) \equiv \overline{R_-^{(2,0)}}\,,\\
        \Lambda_\pm^{(2,l)} & \equiv - \Lambda_R - i \Lambda_I(2 -2l \pm 1)\,,
    \end{split}
\end{equation}
where $\overline{R^{2,l}_\pm}$ denotes the complex conjugate of $R^{2,l}_\pm$, $l = 0,1,2$.

The second step is applying a third order transformation, $\boldsymbol{\phi}_3(\delta X_t^\prime,\delta Y_t^\prime)$, in order to remove redundant third order terms and to obtain a normal form that is equivalent to its polar version provided in Eq. (\ref{eq:CubicNormalFormPolar}). From this third-order transformation the important coefficient is
\begin{equation}\small
\begin{split}
    R_+^{(3,2)} &\equiv \frac 18 \big[ 3 (h_1 R_{1,4} + h_2 R_{2,5}) + h_2 R_{1,4} + h_3 R_{1,5} + h_1 R_{2,5} + h_3 R_{2,4}\big] \\
    & \quad + \frac i8 \big[ 3(h_1 R_{2,4} - h_2 R_{1,4}) + h_2 R_{2,4} + h_3 R_{2,5} - h_1 R_{1,5} - h_3 R_{1,4} \big]\,,
\end{split}
\end{equation}
which determines the coefficient $\mathfrak l$ as
\begin{equation}\small
    \mathfrak l \equiv R_+^{(3,2)} - \phi_+^{(2,1)}(R_+^{(2,2)} + R_-^{(2,1)})-2 (\phi_+^{(2,2)}R_+^{(2,1)} + \phi_+^{(2,0)}R_-^{(2,2)})\,.
\end{equation}
The cubic normal form for a Hopf-bifurcation in polar coordinates is then given by Eq. (\ref{eq:CubicNormalFormPolar}).

\subsection{Back transformation}
Starting in the normal form coordinates (compare Eq. (\ref{eq:LimitCycle})), 
\begin{equation}
    \begin{pmatrix}
        U_t\\ V_t
    \end{pmatrix} = r_{LC} \begin{pmatrix}
        \sin(\omega_{LC}t)\\ \cos(\omega_{LC}t)
    \end{pmatrix}\,,\label{eq:App:LimitCycle}
\end{equation}
we here describe the transformation back to the original coordinates up to the quadratic order. The starting point is the observation that the inverse of the near-identity transformation, Eq. (\ref{eq:App:NearIdentityTransform}), is, to leading order, given by
\begin{equation}
    \boldsymbol \phi_k^{-1}(\delta X_t^\prime,\delta Y_t^\prime) = \begin{pmatrix}
        \delta X_t^\prime\\\delta Y_t^\prime
    \end{pmatrix} -  \boldsymbol \phi^{(k)}(\delta X_t^\prime,\delta Y_t^\prime)\,.\label{eq:App:NearIdentityTransformBack}
\end{equation}
Since the third-order transformation $\boldsymbol \phi_3^{-1}(\delta X_t^\prime,\delta Y_t^\prime)$ introduces cubic corrections to the identity, we can neglect this transformation and directly apply the inverse of the second-order transformation,
\begin{equation}
    \begin{pmatrix}
        \delta \tilde X_t\\ \delta \tilde Y_t
    \end{pmatrix} \equiv \boldsymbol \phi_2^{-1}(U_t,V_t)\,,
\end{equation}
which gives the coordinates on the center manifold up to quadratic order. Thus, the remaining coordinate is given by
\begin{equation}
    \delta \tilde Z_t \equiv  h(\delta \tilde X_t,\delta \tilde Y_t)\,,
\end{equation}
with $h(\delta \tilde X_t,\delta \tilde Y_t)$ as defined in Eq. (\ref{eq:App:hPolynomial}).

The last step is to reverse the basis-transformation into the eigenspace,
\begin{equation}
    \boldsymbol{\delta \alpha}_t \simeq S\boldsymbol{\delta \tilde \alpha}_t\,.
\end{equation}
Since $S$ describes a linear transformation, the correct quadratic order terms are preserved.

\remark{Remark (properties of $\boldsymbol F(U_t,V_t)$):} In the main text, we introduce $\boldsymbol F(U_t,V_t)$ as abbreviation for the different steps discussed above, i.e.,
\begin{equation}
 \boldsymbol{\delta \alpha_t} = \boldsymbol F(U_t,V_t)\,.
\end{equation}
This function is a composition of the near-identity transformation $ \phi_2^{-1}(U_t,V_t)$ and the basis transformation $S$  in two components. The remaining component is the composition of $\phi_2^{-1}(U_t,V_t)$ with the coordinate function on the center manifold, $h(\delta X, \delta Y)$ and the basis transformation $S$. All functions map the origin onto the origin, thus, we have $\boldsymbol F(0,0) = \boldsymbol 0$, or in other words, $\boldsymbol F(U_t,V_t)$ has component functions that are polynomials without any constant term. Furthermore, since $\boldsymbol F(U_t,V_t)$ is a polynomial, collecting the different powers of the limit cycle radius $r_{LC} \propto \sqrt{\Delta S}$ leads to the expansion of $\boldsymbol{\delta \alpha}_t$, see Eq. (\ref{eq:Deviations}) of the main text or Eq. (\ref{eq:app:alphaSeries}) below.

\section{Irreversible approximation}
\label{app:Irreversible}

We consider RRE (\ref{eq:RRE}) and nondimensionalize the concentrations and time as defined in Eq. (\ref{eq:Nondimension}).
We require that the RRE expressed in the re-scaled parameters should still be consistent with a description in terms of a reversible CRN following mass-action dynamics. This leads to
\begin{equation}
    \lambda_X = \lambda_Y = \lambda_Z \equiv \lambda\,.
\end{equation}
In order to stay close to the parameter regime examined in Sec. \ref{sec:Numerics}, we require as many reaction rates of the forward reactions $k_\rho$ as possible to be equal to one. Furthermore, we require that the numerical values of the new kinetic constants should at most be of order one when plugging in the numerical values of Sec. \ref{sec:Numerics}. This leads to re-scaling parameters,
\begin{equation}
    \begin{split}
        \lambda & = \frac{k_2}{k_5}\,,\\
        \lambda_t & = \frac1{k_2}\,,\\
        \lambda_S & = \frac{k_2}{k_1}\,,\\
        \lambda_P & = \frac{k_2^2}{k_5k_{-4}}\,,
    \end{split}
\end{equation}
and the following RRE for the $\mathcal S_i$ species,
\begin{equation}
    \begin{split}
        \frac{d}{dt}\begin{pmatrix}
[X]_t\\
[Y]_t\\
[Z]_t\end{pmatrix} = 
\begin{pmatrix}
1& -1& 0& 0& -1\\
0& 0& 1& -1& 0\\
0& 1& -1& 0& 0
\end{pmatrix}
\begin{pmatrix}
[S][X]_t - \kappa_{-1} [X]_t^2\\
[X]_t - \kappa_{-2} [Z]_t\\
\kappa_3 [Z]_t - \kappa_{-3} [Y]_t\\
\kappa_4 [Y]_t -[P]\\
[X]_t[Y]_t - \kappa_{-5} [P][Y]_t
\end{pmatrix}\,,
    \end{split}
\end{equation}
where we dropped the $\tilde {}$-symbol for convenience. The new "kinetic constants" are defined as
\begin{equation}
\begin{split}
\kappa_{-1} &\equiv k_{-1}\lambda_t \lambda = \frac{k_{-1}}{k_5}  = 10^{-2}\,,\\
\kappa_{-2} &\equiv k_{-2}\lambda_t = \frac{k_{-2}}{k_2}= 10^{-2}\,,\\
\kappa_3 &\equiv k_3\lambda_t = \frac{k_3}{k_2}= 1\,,\\ 
\kappa_{-3} &\equiv k_{-3} \lambda_t = \frac{k_{-3}}{k_2}  = 10^{-2}\,,\\
\kappa_4 &\equiv  k_4\lambda_t = \frac{k_4}{k_2} = 1\,,\\
\kappa_{-5} &\equiv k_{-5}\lambda_P\lambda_t = \frac{k_2}{k_5}\frac{k_{-5}}{k_{-4}} = 10^{-4}\,,
\end{split}\label{eq:NewConstants}
\end{equation}
with their numerical value corresponding to Sec. \ref{sec:Numerics}.

Based on the numerical values of the constants in Eq. (\ref{eq:NewConstants}), we define the irreversibility limit as
\begin{equation}
    \kappa_{-1},\kappa_{-2},\kappa_{-3}, \kappa_{-5} \ll 1\,.
\end{equation}
Note that $\kappa_{-4}$ remains finite in this limit, while the remaining backward rate constants become negligible.
In this limit, CRN (\ref{eq:CRN}) is approximated by the CRN (\ref{eq:CRNirr}) in the main text.

\section{Approach to the irreversible limit}
\label{app:irr_convergence}

To examine the quality of the irreversible approximation, we introduce a one-parameter family of reversible CRNs in which the backward reaction rates scale as
\begin{equation}
    \kappa_{-1}=\kappa_{-2}=\kappa_{-3}=\varepsilon,\qquad \kappa_{-5} =\varepsilon^2,
\end{equation}
while the Wegscheider condition, Eq. (\ref{eq:Wegscheider}), in the re-scaled rates,
\begin{equation}
	\frac{\kappa_{3} \kappa_{4} \kappa_{-5}}{\kappa_{-2} \kappa_{-3}} = 1\,, \label{eq:App:Wegscheider0}
\end{equation}
is preserved. The parameter $\varepsilon$ therefore controls the distance from the irreversible limit $\varepsilon\to0$.

\begin{figure}
    \centering
    \includegraphics[width=\figwidth]{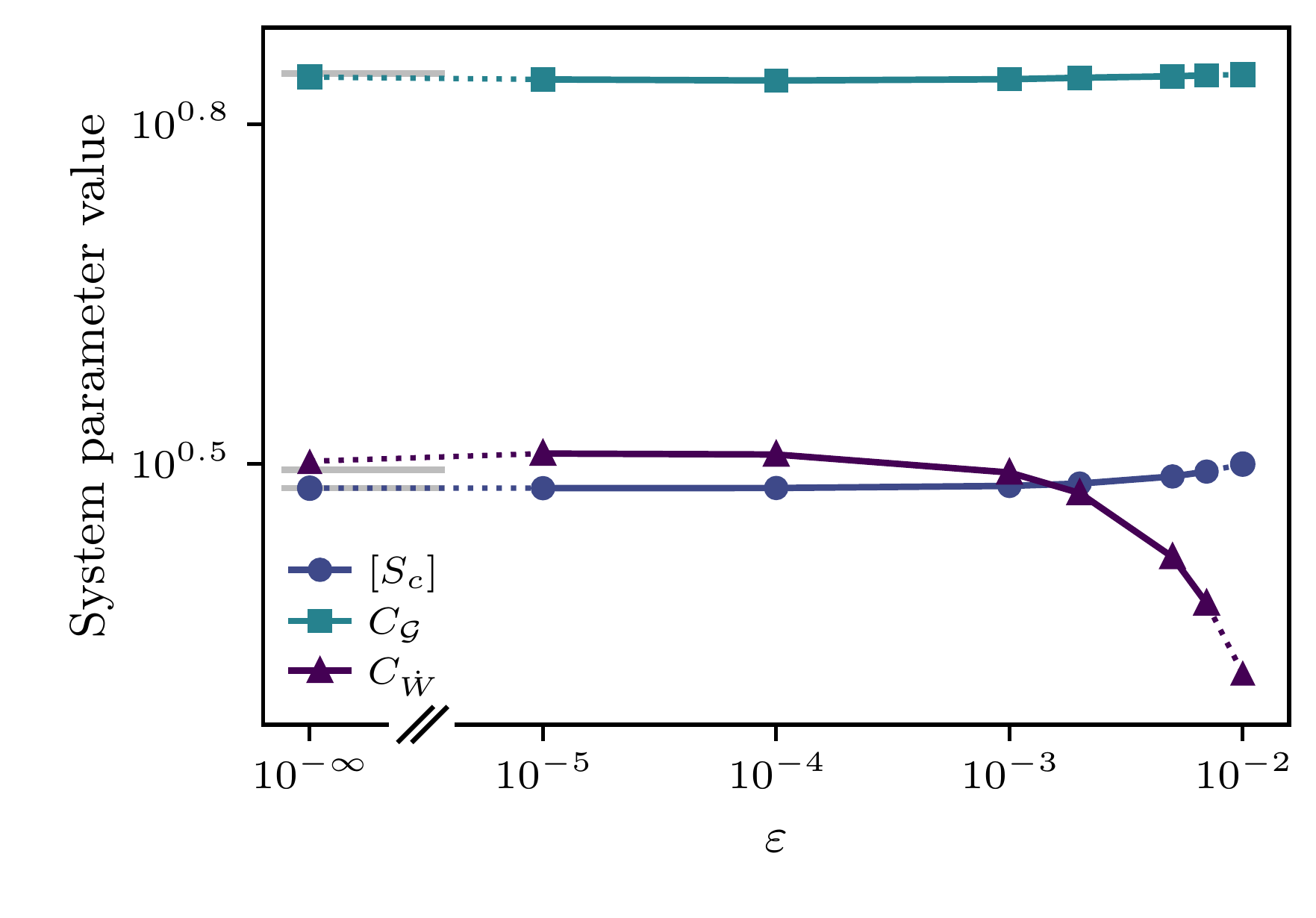}
    \caption{
Approach to the irreversible limit.
Backward rates are parameterized by $\kappa_{-1}=\kappa_{-2}=\kappa_{-3}=\varepsilon$ and $\kappa_{-5} =\varepsilon^2$, preserving the Wegscheider condition. Shown are the numerically extracted Hopf threshold $[S_c]$ and jump coefficients $C_{\mathcal G}$ and $C_{\dot W}$ as functions of $\varepsilon$.  Horizontal lines indicate the analytical predictions from the irreversible approximation ($\varepsilon\to0$). The convergence demonstrates that the irreversible model captures the leading-order behavior of the reversible CRN.
}
    \label{fig:App:IrrLimit}
\end{figure}

For each value of $\varepsilon$, we numerically determine the Hopf threshold $[S_c]$ and the jump coefficients $C_{\mathcal G}$ and $C_{\dot W}$ defined in Eqs.~(\ref{eq:Gdiff}) and (\ref{eq:Wdiff}) by a linear fit. The results are shown in Fig.~\ref{fig:App:IrrLimit}.

We find that the critical value $[S_c]$ and the coefficient $C_{\mathcal G}$ rapidly converge toward the analytical prediction obtained in the irreversible approximation. The coefficient $C_{\dot W}$ shows larger quantitative corrections for finite $\varepsilon$, but approaches the same limiting value as $\varepsilon\to0$. This supports the use of the irreversible CRN as a controlled asymptotic approximation of the reversible dynamics in the regime of weak backward reactions.

\begin{figure}[t]
\centering
   \includegraphics[width=\figwidth]{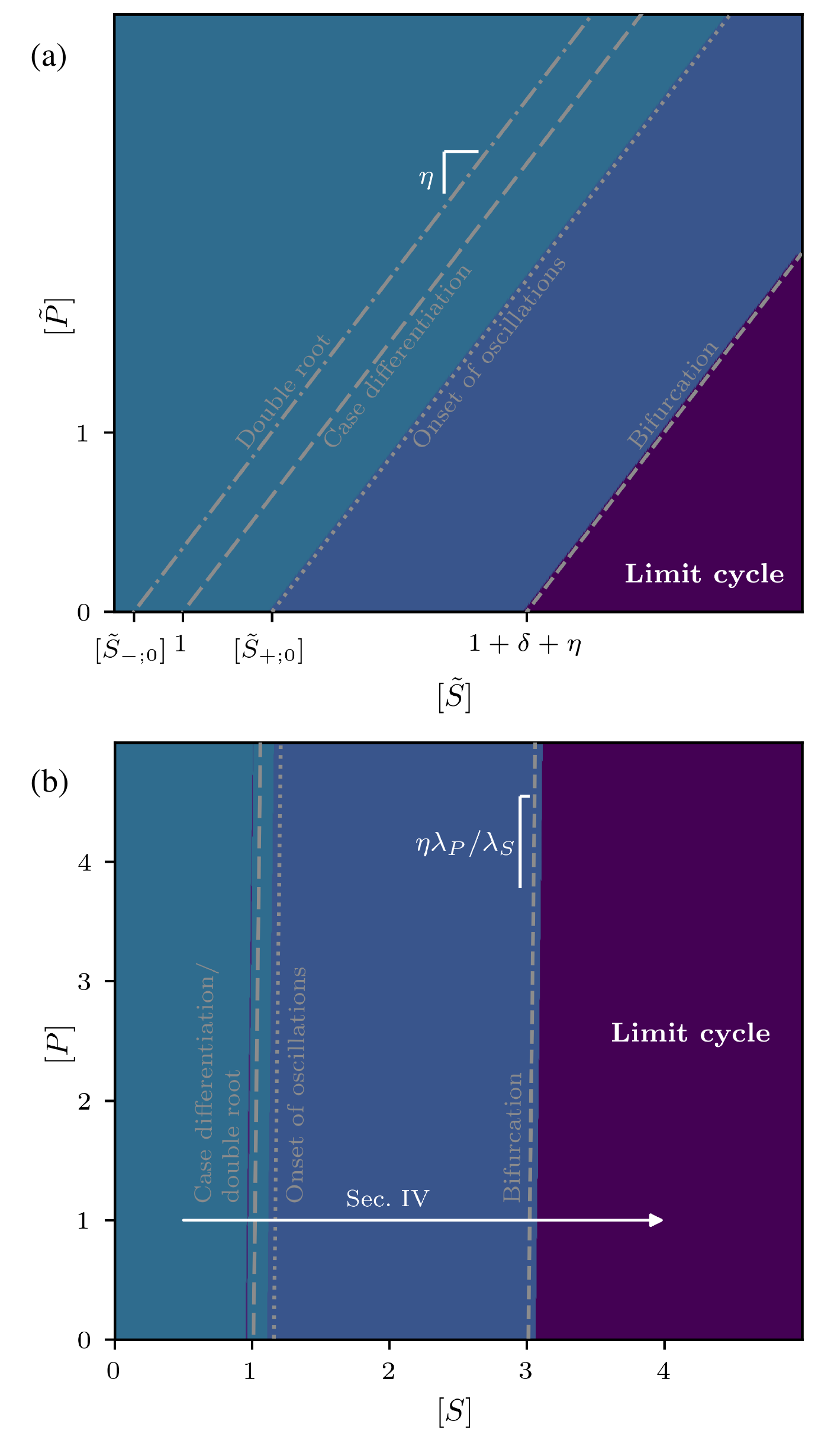}
\caption{ Phase space of the irreversible and reversible CRNs.
(a) Schematic phase-space sketch of the irreversible CRN (\ref{eq:CRNirr}) using dimensionless concentrations, Eq.~(\ref{eq:Nondimension}). The boundaries $[\tilde S_{\pm;0}] = \frac{- 2 \delta^3 + 27 \delta \eta + 3 \delta^2 \eta + 3 \delta \eta^2 - 2 \eta^3 \pm 2 (\delta^2 - \delta \eta + \eta^2)^{3/2}}{27 \delta \eta}$ are obtained by setting $[\tilde P] = 0$ in Eq. (\ref{eq:OnsetOfOsci}).
(b) Numerical comparison of the phase space of the reversible CRN (\ref{eq:CRN}) and its irreversible approximation, CRN (\ref{eq:CRNirr}), in physical concentrations. Gray dashed lines represent the results of the irreversible approximation, while the contour plot shows data from the full reversible CRN. Colors highlight the different regimes, while phase-boundaries correspond to the grey lines in the irreversible approximation.
Parameters (arbitrary units): $k_1 = k_2 = k_3 = k_4 = k_5 = 1$, $k_{-1} = k_{-2} = k_{-3} = k_{-4} = 10^{-2}$, and $k_{-5} = 10^{-6}$.
} 
    \label{fig:Phase}
\end{figure}

\section{Linear stability analysis}
\label{app:LinearStability}

We here determine the fixed points of the irreversible CRN (\ref{eq:CRNirr}) as functions of the control parameters $[S]$ and $[P]$. In the irreversible approximation, the RRE for the $\mathcal S_i$ species reads
\begin{equation}
    \frac{d}{dt}\begin{pmatrix}
[X]_t\\
[Y]_t\\
[Z]_t\end{pmatrix} = \begin{pmatrix}
[X]_t([S] - 1 - [Y]_t)\\
\delta [Z]_t - \eta [Y]_t + [P]\\
[X]_t -  \delta[Z]_t
\end{pmatrix} \,,\label{app:eq:RREirr}
\end{equation}
compare Eq. (\ref{eq:RREirr}) in the main text.
Setting the right-hand-side of Eq. (\ref{app:eq:RREirr}) to zero yields two solutions. For $[S] \geq[S_\mathrm{case}] \equiv 1 + [P]/\eta$, the fixed points are
\begin{equation}
    \begin{split}
        [Y_*] &= [S]-1\,,\\
        [Z_*] &= \frac{\eta[Y_*]-[P]}{\delta} = \frac{\eta}\delta ([S]-1) - \frac {[P]}\delta\,,\\
        [X_*] &= \delta[Z_*] =   \eta([S]-1) - [P]\,.
    \end{split}\label{eq:FP1}
\end{equation}
for $[S] <[S_\mathrm{case}]$, 
\begin{equation}
    \begin{split}
        [Y_*] &= \frac{[P]}\eta\,,\\
        [X_*] &= [Z_*] =  0\,,
    \end{split}\label{eq:FP2}
\end{equation}
with a smooth transition from one solution, Eq. (\ref{eq:FP1}), to the other, Eq. (\ref{eq:FP2}).

\subsection{Jacobian}
\label{app:Eigenvalues}

\subsubsection{Set-up}
The linear stability analysis of the irreversible RRE (\ref{eq:RREirr}) relies on the eigenvalues of the Jacobian matrix evaluated in the fixed point. Here, we calculate the Jacobian matrix and provide the corresponding eigenvalues in the appropriate parameter regimes.

The Jacobian matrix corresponding to the irreversible RRE (\ref{eq:RREirr}) is given by
\begin{equation}
\small\begin{split}
    J &\equiv \begin{pmatrix}
        \partial_{[X]_t} \big(\frac{d}{dt}[X]_t \big) & \partial_{[Y]_t} \big(\frac{d}{dt}[X]_t \big) & \partial_{[Z]_t} \big(\frac{d}{dt}[X]_t \big)\\
        \partial_{[X]_t} \big(\frac{d}{dt}[Y]_t \big) & \partial_{[Y]_t} \big(\frac{d}{dt}[Y]_t \big) & \partial_{[Z]_t} \big(\frac{d}{dt}[Y]_t \big)\\
        \partial_{[X]_t} \big(\frac{d}{dt}[Z]_t \big) & \partial_{[Y]_t} \big(\frac{d}{dt}[Z]_t \big) & \partial_{[Z]_t} \big(\frac{d}{dt}[Z]_t \big)
    \end{pmatrix}\bigg|_{[X_*],[Y_*],[Z_*]}\\ 
    &  = \begin{pmatrix}
        [S]-1 - [Y_*]& -[X_*]  &0 \\
        0          & -\eta &\delta \\
        1          & 0     &-\delta
    \end{pmatrix}\,.
\end{split}\label{eq:Jacobian}
\end{equation}

\subsubsection{Case: $[S]< [S_\mathrm{case}]$}
In the parameter range, $[S] <[S_\mathrm{case}]$, the fixed points are given in Eq. (\ref{eq:FP2}). Plugging Eq. (\ref{eq:FP2}) into Eq. (\ref{eq:Jacobian}) yields
\begin{equation}
    J_< \equiv \begin{pmatrix}
[S]-[S_\mathrm{case}]& -0& 0 \\
0 & - \eta& \delta\\
1 & 0 & - \delta
\end{pmatrix}\,.
\end{equation}
The associated characteristic polynomial reads
\begin{equation}
\begin{split}
    \chi_{J_<}(\Lambda)& \equiv \det (J_< - \Lambda \mathbb 1)\\
    &= -(\Lambda - ([S]-[S_\mathrm{case}]))(\Lambda + \eta)(\Lambda + \delta)\,,\label{eq:CharacPolBelow}
\end{split}
\end{equation}
where $\det$ denotes the determinant and $\mathbb 1$ the $3\times 3$ identity-matrix. The roots of $\chi_{J_<}(\Lambda)$ given in Eq. (\ref{eq:CharacPolBelow}) determine the eigenvalues,
\begin{equation}
    \begin{split}
        \Lambda_0 &\equiv -([S_\mathrm{case}]-[S])\,,\\
        \Lambda_1 &\equiv -\eta\,,\\
        \Lambda_2 &\equiv -\delta\,,
    \end{split}\label{eq:EigenvaluesBelow}
\end{equation}
which are negative for $[S]<[S_\mathrm{case}]$. Thus, the fixed points given in Eq. (\ref{eq:FP2}) are stable and no Hopf bifurcation can occur in this parameter range.

\subsubsection{Case: $[S] \geq [S_\mathrm{case}]$}
In the parameter-regime above $[S_\mathrm{case}]$, the fixed points are given in Eq. (\ref{eq:FP1}). Plugging the fixed points into Eq. (\ref{eq:Jacobian}), the Jacobian matrix reads
\begin{equation}
    J_> \equiv \begin{pmatrix}
0 & [P]-\eta ([S]-1)& 0 \\
0 & - \eta& \delta\\
1 & 0 & - \delta
\end{pmatrix}\,,
\end{equation}
with associated characteristic polynomial
\begin{equation}
    \chi_{J_>}(\Lambda)= -\Lambda^3 - (\eta+\delta)\Lambda^2 - \eta \delta \Lambda - \delta \eta ([S]-1) + \delta [P]\,.\label{eq:CharacPolyAbove}
\end{equation}

Using Mathematica, the roots of Eq. (\ref{eq:CharacPolyAbove}) are given by
\small{
\begin{equation}
\begin{split}
\Lambda_0 &= \frac{1}{12}\left(
-4(\delta+\eta)
+ \frac{2\,2^{1/3}(1+i\sqrt{3})A}{D}
+ 2^{2/3}(1-i\sqrt{3})D
\right)\,,\\
\Lambda_1 &= \frac{1}{12}\left(
-4(\delta+\eta)
+ \frac{2\,2^{1/3}(1-i\sqrt{3})A}{D}
+ 2^{2/3}(1+i\sqrt{3})D
\right)\,,\\
\Lambda_2 &= \frac{1}{6}\left(
-2(\delta+\eta) - \frac{2\,2^{1/3}A}{D} + 2^{2/3}D
\right)\,,
\end{split}\label{eq:Eigenvalues}
\end{equation}
}
with the following abbreviations
\small{
\begin{equation}
\begin{split}
A &\equiv \delta^2 - \delta\eta + \eta^2\,,\\
B &\equiv -27[P]\delta + 2\delta^3 - 27\delta\eta + 27[S]\delta\eta
      - 3\delta^2\eta - 3\delta\eta^2 + 2\eta^3\,,\\
C &\equiv 27[P]\delta - 2\delta^3 + 3\delta^2\eta - 2\eta^3
      + 3\delta\eta(9-9[S]+\eta)\,,
\end{split}
\end{equation}
}
and
\begin{equation}
\begin{split}
\Delta &\equiv \sqrt{-4A^3 + C^2}\,,\label{eq:Discriminant}\\
D&\equiv \bigl(B+\Delta\bigr)^{1/3}\,.
\end{split}
\end{equation}

\subsection{Onset of decaying oscillations}
\label{Sec:OnsetOfOsci}

The Hopf line is discussed in the Sec. \ref{sec:HopfIrr}. Here, we determine the line in phase space where decaying oscillations emerge. In order to do so, we consider the discriminant $\Delta$ of the characteristic polynomial $\chi_{J_>}(\Lambda)$ as defined in Eq. (\ref{eq:Discriminant}).
The discriminant $\Delta$ of a cubic polynomial vanishes if and only if the polynomial has a multiple root.

Using Mathematica, the parameterized roots of $\Delta$ read
\begin{equation}\small
    [S_\pm] \equiv \frac{27[P]\delta - 2 \delta^3 + 27 \delta \eta + 3 \delta^2 \eta + 3 \delta \eta^2 - 2 \eta^3 \pm 2 (\delta^2 - \delta \eta + \eta^2)^{3/2}}{27 \delta \eta}\,.\label{eq:OnsetOfOsci}
\end{equation}
For the parameter-set considered in Sec. \ref{sec:Numerics}, the solutions $[S_\pm]$ become
\begin{equation}
    [S_-] = 1.01 = [S_\mathrm{case}]\,, \quad [S_\mathrm{osc}] \equiv [S_+] = \frac{3127}{2700}\approx 1.158\,.
\end{equation}
Thus, we identify $[S_-]$ with a double root touching the abscissa, and $[S_\mathrm{osc}] \equiv [S_+]$ as the onset of oscillations where, after crossing this root, the characteristic polynomial develops a pair of complex conjugate roots.

\remark{Remark:} For the parameter-set considered in Sec. \ref{sec:Numerics}, the characteristic polynomial has a double root at $[S_-] = [S_\mathrm{case}]$ since two eigenvalues coincide: $\Lambda_1 = -\eta = -1 = -\delta = \Lambda_2$.

\subsection{Map of the phase space}
\label{app:PhaseSpace}

The phase-space structure derived in the previous sections and in Sec. \ref{sec:HopfIrr} is illustrated in Fig. \ref{fig:Phase}. The general structure is shown in Fig. \ref{fig:Phase}a, where the double root and the different cases can be clearly distinguished.

We determine the phase-space structure of the full (reversible) CRN (\ref{eq:CRN}) numerically by performing a linear stability analysis and evaluating the Hopf condition, Eq. (\ref{eq:HopfCondition}). Comparing these numerical results with the analytical predictions for the irreversible CRN (\ref{eq:CRNirr}), we find good agreement, see Fig. \ref{fig:Phase}b. In particular, key features of the phase-space geometry, such as the number and slope of the phase boundaries, are well reproduced by the irreversible approximation. However, while the location of the phase boundaries is captured within a reasonable range, quantitative deviations remain (see also Sec. \ref{sec:IrreversibleSummary}). 
For the parameter set considered here, the lines corresponding to the double root and case differentiation coincide.

\section{Expressions obtained with Mathematica}
\label{app:MathematicaResults} 
We here state results from the accompanying Mathematica notebook that are supporting the results obtained in the main text for the irreversible CRN (\ref{eq:CRNirr}).

\remark{Limit cycle radius and angular frequency:} The limit cycle radius is
\begin{equation}
    r_{LC} \equiv \sqrt{- \frac{\Lambda_R}{\Re(\mathfrak l)}}  = \sqrt{2\frac{\eta\big((\eta + \delta)^2 + 4 \eta \delta \big)}{\delta^2}}\sqrt{\Delta S}\label{app:eq:rLC}
\end{equation}
The angular frequency of the limit cycle is given by
\begin{widetext}
\begin{equation}
\omega_{LC} \equiv \Lambda_I  + \Im (\mathfrak l)\frac{\Lambda_R}{\Re(\mathfrak l)} = \sqrt{\eta \delta} + \Delta S \frac{-\delta^7 - 3 \delta^6\eta + 4 \delta^5 \eta^2 + 18 \delta^{9/2}\eta^{5/2}+20 \delta^4 \eta^3 + 2 \delta^3\eta^4 + 4 \delta^2 \eta^5 - \delta \eta^6 - \eta^7 + 36 (\delta \eta)^{7/2}}{12 \sqrt{\eta \delta}(\eta + \delta)^2 (\eta^2 + 3 \delta \eta+ \delta^2)^3}\,.\label{app:eq:omegaLC}
\end{equation}
\end{widetext}

\remark{Expansion of $\boldsymbol{\delta \alpha}_t$:}
The deviations from the fixed point $[\boldsymbol \alpha_*]$ can be expanded in the control parameters $\Delta S$ as follows
\begin{align}
\boldsymbol{\delta \alpha}_t
&=
\boldsymbol{c}^{(1/2)}_t\,\sqrt{\Delta S}
+\boldsymbol{c}^{(1)}_t\,\Delta S
+\mathcal{O}\!\big((\Delta S)^{3/2}\big)\,,\label{eq:app:alphaSeries}
\end{align}
compare Eq. (\ref{eq:Deviations}). The expansion coefficients are
\begin{align}
\boldsymbol{c}^{(1/2)}_t
=
\sqrt{2}\,\sqrt{\eta M_0}
\begin{pmatrix}
\cos t+\sqrt{\tfrac{\eta}{\delta}}\sin t\\
\dfrac{1}{\delta+\eta}
\Bigl(\cos t-\sqrt{\tfrac{\delta}{\eta}}\sin t\Bigr)\\
\dfrac{1}{\delta}\cos t
\end{pmatrix}\,,\label{eq:App:c12}
\end{align}
with 
\begin{equation}
    M_0 \equiv \delta^2+6\delta\eta+\eta^2\,,
\end{equation}
and
\begin{align}
\boldsymbol{c}^{(1)}_t
=
\frac{2}{(\delta+\eta)M_1}
\begin{pmatrix}
\sqrt{\tfrac{\eta}{\delta}}\,\{ M_2(7\delta+5\eta)\cos(2t)+M_3\sin(2t)\}\\
\dfrac{1}{\sqrt{\delta\eta}}\,\{M_2(7\delta+5\eta)\cos(2t)+M_3\sin(2t)\}\\
\displaystyle
\frac{\sqrt{\eta}}{\delta}\,M_4\cos(2t)
+\frac{\sqrt{\eta}}{2\delta^{3/2}}\,M_5\sin(2t)
\end{pmatrix}\,,\label{eq:App:c1}
\end{align}
where
\begin{equation}
\begin{split}
M_1 &\equiv \delta^2+3\delta\eta+\eta^2\,,\\
M_2 &\equiv \sqrt{\delta^7\eta}+\sqrt{\delta\eta^7}+(\delta\eta)^{3/2}\,,\\
M_3 &\equiv -\delta^4-5\delta^3\eta+6\delta^2\eta^2+5\delta\eta^3+\eta^4\,,\\
M_4 &\equiv 2\delta^3+15\delta^2\eta+8\delta\eta^2+\eta^3\,,\\
M_5 &\equiv -3\delta^4-15\delta^3\eta+6\delta^2\eta^2+7\delta\eta^3+\eta^4\,.
\end{split}
\end{equation}
Using $\int_0^{2\pi}\cos tdt = \int_0^{2\pi}\sin tdt = \int_0^{2\pi}\cos 2tdt = \int_0^{2\pi}\sin 2tdt = 0$ in Eq. (\ref{eq:App:c12}) and (\ref{eq:App:c1}) yields $\int_0^{2\pi} \boldsymbol c^{(1/2)}_t dt = \int_0^{2\pi} \boldsymbol c^{(1)}_t dt = \boldsymbol 0$.

\remark{Expansion of $\braket{\mathcal G} - \mathcal G_*$:}

We report the expansion coefficient of the semi-grand Gibbs free energy around the bifurcation as explained in Sec. \ref{sec:Thermo}, which reads
\begin{align}
\braket{\mathcal G}- \mathcal G_*
=
\begin{cases}
C_{\mathcal G}\Delta S\,,
& \Delta S \ge 0\,,
\\
0\,,
& \Delta S < 0\,,
\end{cases}
\end{align}
with
\begin{equation}\small
    C_\mathcal G \equiv
\frac{(\delta^2+6\delta\eta+\eta^2)\Bigl([P](1+\delta+\eta)+\eta(\delta^2 + 2 \delta(1+\eta) + \eta(1+\eta))\Bigr)}
{2\delta(\delta+\eta)\bigl([P]+\eta(\delta+\eta)\bigr)}.\label{eq:App:CG}
\end{equation}

\remark{Expansion of $\braket{\dot W} - \dot W_*$:}
Performing the expansion of the work rate around the bifurcation as explained in Sec. \ref{sec:Thermo} yields
\begin{align}
\braket{\dot W} - \dot W_*
=
\begin{cases}
-C_{\dot W}\Delta S\,,
& \Delta S \ge 0\,,
\\
0\,,
& \Delta S < 0\,,
\end{cases}
\end{align}
where
\begin{equation}\small
    C_{\dot W} \equiv
\frac{\kappa_{-1}\,\eta(\delta+\eta)(\delta^2+6\delta\eta+\eta^2)}{\delta}
\log\!\left(
\frac
{[P]+\eta(1+\delta+\eta)}{\kappa_{-1}\,\kappa_{-5}\,[P]\,\eta}
\right)\,.\label{eq:App:CW}
\end{equation}

\section{Second derivative of the work rate}
\label{app:WorkRate}

We derive here Eq. (\ref{eq:RateCondition}) of the main text and discuss its implications in the irreversibility limit.

The second derivative of the average work rate at the Hopf-bifurcation evaluated in the fixed point reads
\begin{equation}
	 \partial^2_{[X]}\dot W([X_*]) = - 2\kappa_{-1} (\ln [S_{\mathrm{HB}}] - \ln [P] + \mu_S^\circ - \mu_P^\circ)\,,\label{eq:App:WorkRateSecondDerivative}
\end{equation}
compare Eq. (\ref{eq:WorkRateSecondDerivative}).
The kinetic constant $\kappa_{-1}$ is positive, hence, the sign of  $\partial^2_{[X]}\dot W([X_*])$ is determined by the sign of the term in brackets on the right hand side of Eq. (\ref{eq:App:WorkRateSecondDerivative}). Instead of treating this expression directly, we examine the following term,
\begin{equation}
	\mu([S]) - \mu([P]) = \ln [S] - \ln [P] + \mu_S^\circ - \mu_P^\circ\,,\label{eq:App:WorkRateTerm0}
\end{equation}
which corresponds to the thermodynamic driving between the chemostats $S$ and $P$, and determine in which parameter-regime it has a positive sign.

\begin{figure}
\centering
		\includegraphics[width=\figwidth]{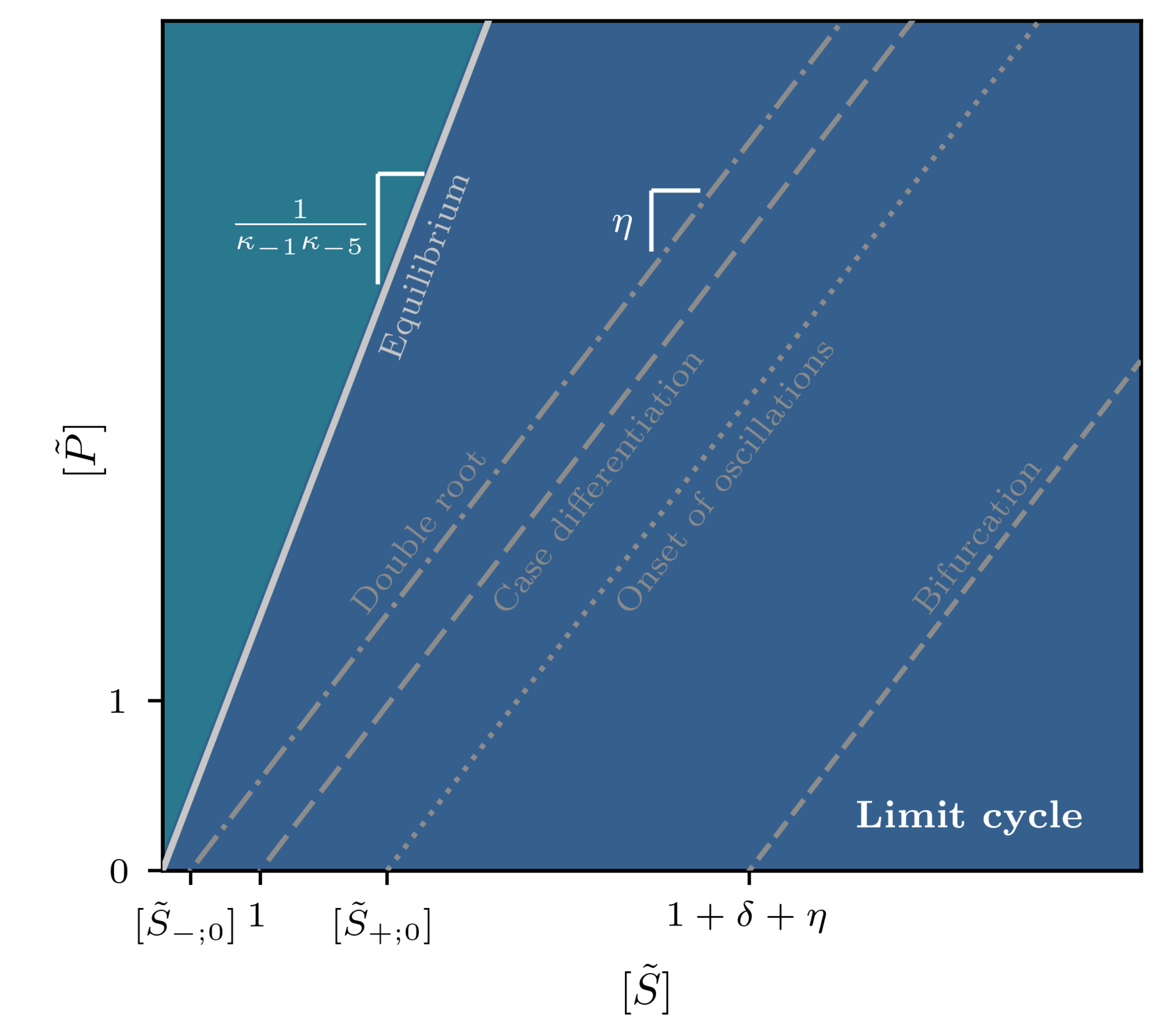}
\caption{ Phase-space sketch of the irreversible CRN (\ref{eq:CRNirr}) including the equilibrium line, Eq. (\ref{eq:App:RateCondition}).
The blue shaded region indicates $\mu([S]) - \mu([P]) > 0$ between the chemostats, Eq. (\ref{eq:App:WorkRateTerm}).} 
    \label{fig:App:Phase}
\end{figure}

Local detailed balance, Eq. (\ref{eq:LDB}), gives
\begin{equation}
	\mu_S^\circ - \mu_P^\circ = -\ln \kappa_{-1} - \ln \frac{\kappa_{-2} \kappa_{-3}}{\kappa_{3} \kappa_{4}}\,.\label{eq:App:WorkRateTermMu0}
\end{equation}
The CRN discussed in Sec. \ref{sec:Numerics} satisfies the Wegscheider condition for the internal cycle $\boldsymbol c$, compare Eq. (\ref{eq:Wegscheider}),
which, expressed in the re-scaled kinetic constants, reads,
\begin{equation}
	\frac{\kappa_{3} \kappa_{4} \kappa_{-5}}{\kappa_{-2} \kappa_{-3}} = 1\,. \label{eq:App:Wegscheider}
\end{equation}
With Eqs. (\ref{eq:App:Wegscheider}) and (\ref{eq:App:WorkRateTermMu0}), Eq. (\ref{eq:App:WorkRateTerm0}) becomes
\begin{equation}
	\mu([S]) - \mu([P])= \ln \frac{[S]}{[P]\kappa_{-1}\kappa_{-5}}\,,\label{eq:App:WorkRateTerm}
\end{equation}
which has a positive sign if
\begin{equation}
	 \frac{[S]}{\kappa_{-1}\kappa_{-5}} >[P]\,.
\end{equation}
Geometrically, in the $[S]-[P]$-parameter-space, Eq. (\ref{eq:App:WorkRateTerm}) is positive in the area below the line through the origin with slope $1/\kappa_{-1}\kappa_{-5}$,
\begin{equation}
	[P] =  \frac{[S]}{\kappa_{-1}\kappa_{-5}}\,,\label{eq:App:EquilibriumLine}
\end{equation}
see Fig. \ref{fig:App:Phase}. This line through the origin corresponds to equilibrium, since the thermodynamic driving between the chemostats $S$ and $P$, Eq. (\ref{eq:App:WorkRateTerm}), vanishes. Thus, in the irreversible limit, physically meaningful solutions require that the equilibrium line in phase space and the parameterized Hopf bifurcation line (see Eq. (\ref{eq:HopfLine})),
\begin{equation}
[P] = \eta ([S] - 1 - \eta - \delta)\,,\label{eq:App:HopfLine}
\end{equation}
do not intersect.
Eq. (\ref{eq:App:HopfLine}) parameterizes a line in the $[S]-[P]$-plane, which intersects the abscissa at a positive value. Hence, the lines defined in Eq. (\ref{eq:App:EquilibriumLine}) and Eq. (\ref{eq:App:HopfLine}) do not intersect, as long as their slopes satisfy
\begin{equation}
 \frac{1}{\kappa_{-1}\kappa_{-5}} > \eta\,,\label{eq:App:RateCondition}
\end{equation}
which is Eq. (\ref{eq:RateCondition}) of the main text.

\remark{Remark:} We exclude the equality in Eq. (\ref{eq:App:RateCondition}), for which the lines, Eq. (\ref{eq:App:EquilibriumLine}) and (\ref{eq:App:HopfLine}), would be parallel, since we ultimately consider the limits $\kappa_{-1},\kappa_{-5} \to 0$ while keeping $\eta$ finite.

\remark{Remark:} Physically, Eq. (\ref{eq:App:RateCondition}) ensures that the CRN is driven from chemostat $S$ to $P$ consistent with the direction of current $I_S$.

\typeout{get arXiv to do 4 passes: Label(s) may have changed. Rerun}
\end{document}